\documentclass[12pt]{article}
\usepackage{amsmath,amssymb,graphicx,slashbox}
\usepackage{color}
\usepackage{comment}
\def\mydate{8 August 2011}
\def\ignore#1{{}}

\newcommand{\bea}{\begin{eqnarray}}
\newcommand{\eea}{\end{eqnarray}}

\makeatletter
\@addtoreset{equation}{section}
\makeatother

\newcommand{\simg}{%
\hspace{0.3em}\raisebox{0.4ex}{$>$}\hspace{-0.75em}\raisebox{-.7ex}{$\sim$}\hspace{0.3em}} 

\newcommand{\siml}{%
\hspace{0.3em}\raisebox{0.4ex}{$<$}\hspace{-0.75em}\raisebox{-.7ex}{$\sim$}\hspace{0.3em}}

\newcommand{\beeq}{\begin{equation}}
\newcommand{\eneq}{\end{equation}}
\newcommand{\beqn}{\begin{eqnarray}}
\newcommand{\eeqn}{\end{eqnarray}}

\def\mybig{\displaystyle \strut }

\def\dd{\partial}
\def\la{\raise.16ex\hbox{$\langle$}\lower.16ex\hbox{}  }
\def\ra{\raise.16ex\hbox{$\rangle$}\lower.16ex\hbox{} }
\def\go{\rightarrow}

\def\onehalf{ \hbox{$\frac{1}{2}$} }
\def\onethird{ \hbox{$\frac{1}{3}$} }
\def\twothird{ \hbox{$\frac{2}{3}$} }

\def\tr{{\rm tr \,}}
\def\eff{{\rm eff}}

\def\EM{{\rm EM}}
\def\diag{{\rm diag ~}}

\def\KK{{\rm KK}}
\def\bulk{{\rm bulk}}
\def\brane{{\rm brane}}

\def\psibar{ \psi \kern-.65em\raise.6em\hbox{$-$} }
\def\psibarl{ \psi \kern-.65em\raise.6em\hbox{$-$} \lower.6em\hbox{} }

\def\myeq{\!\!\!=\!\!\!}
\def\myfrac#1#2{{\mybig #1\over \mybig #2}}

\begin{document}

\thispagestyle{empty}

%\begin{titlepage}
%%%%% PREPRINT NUMBERS %%%%%%
{\small \noindent \mydate    \hfill OU-HET 692/2011}

%{\small \noindent \myred{updated YH}    \hfill}

\vspace{4.0cm}

%%%%%%%%%%%%%%%%%%% TITLE %%%%%%%%%%%%%%%%%%
\baselineskip=35pt plus 1pt minus 1pt

\begin{center}
{\LARGE \bf 
Collider signatures of\\
the SO(5)$\times $U(1) gauge-Higgs unification}
\end{center}
%%%%%%%%%%%%%%%% AUTHORS %%%%%%%%%%%%%%%%%%%%%%%

\vspace{2.0cm}
\baselineskip=20pt plus 1pt minus 1pt

\begin{center}
{\bf
Yutaka Hosotani, Minoru Tanaka and Nobuhiro Uekusa
}

%%%%%%%%%%%% AFFILIATION %%%%%%%%%%%%

{\small \it Department of Physics, 
Osaka University, 
Toyonaka, Osaka 560-0043, 
Japan} \\
\end{center}

%%%%%%%%%%% ABSTRACT %%%%%%%%%%%%%%%

\vskip 2.5cm
\baselineskip=20pt plus 1pt minus 1pt

\begin{abstract}
Collider signatures of  the $SO(5) \times U(1)$ gauge-Higgs unification model 
in the Randall-Sundrum warped space are explored.
Gauge couplings of quarks and leptons receive small corrections from the fifth
dimension whose effects are tested by the precision data.  
It is found that the forward-backward asymmetries
in $e^+e^-$  collisions on the $Z$ pole 
are well explained in a wide range of the warp factor $z_L$, 
but the model is consistent with the branching fractions of $Z$ decay
only for large $z_L \simg 10^{15}$.
Kaluza-Klein (KK) spectra of gauge bosons, quarks, and leptons as well as
gauge  and Higgs couplings of low-lying KK excited states are
determined.  Right-handed quarks and leptons have larger couplings to
the KK gauge bosons than left-handed ones.  
Production rates of Higgs bosons and KK states at Tevatron, LHC and  ILC  are evaluated.  
The first KK $Z$ has a mass 1130$\,$GeV with a width 422$\,$GeV for $z_L = 10^{15}$. 
The current limit on the $Z'$ production at Tevatron and LHC indicates $z_L >10^{15}$.
A large effect of parity violation appears in the difference between the rapidity distributions of
$e^+$ and $e^-$ in the decay of the first KK $Z$.
The first KK gauge bosons decay into light and heavy quarks evenly.
\end{abstract}

%\end{titlepage}

%\tableofcontents

\newpage

\baselineskip=20pt plus 1pt minus 1pt

\section{Introduction}
One of the biggest issues in current physics is to
find the Higgs boson and pin down its properties.  
The mechanism of electroweak (EW) symmetry breaking is yet to be
uncovered.  It is not clear if the EW symmetry is spontaneously
broken in a way described in the standard model (SM). 
The search for the Higgs boson is carried on at Tevatron and LHC.
The forthcoming result will tell us whether or not the SM scenario of 
the Higgs boson with a mass $< 200\,$GeV is correct.

Many alternative  models have been proposed
with new physics beyond the standard model.
The most popular scenario in this category is supersymmetry 
hidden at the TeV scale.   The Higgs boson is absent in the Higgsless 
model in which Kaluza-Klein (KK) modes of gauge bosons in the fifth
dimension unitarize the theory,\cite{Csaki1, Cacciapaglia1} 
whereas the Higgs boson appears as a pseudo-Nambu-Goldstone boson
 in the little Higgs model.\cite{Arkani1, Kaplan1, Schmaltz1}
In the gauge-Higgs unification scenario the Higgs boson is unified with
4D gauge fields, appearing as a part of the extra-dimensional component
of gauge potentials.

In the present paper we focus on phenomenological implications and 
predictions of gauge-Higgs unification,\cite{YH1}-\cite{Csaki2}
particularly of the $SO(5) \times U(1)$
model in the Randall-Sundrum (RS) warped space.\cite{ACP}-\cite{HNU}
The Higgs boson is nothing but a four-dimensional fluctuation mode of the 
Wilson line phase $\theta_H$ representing an Aharonov-Bohm phase 
in the fifth dimension.
It has been shown  in a class of the $SO(5) \times U(1)$ models
that the energy density is minimized at $\theta_H = \pm \onehalf \pi$ \cite{HOOS}
where the Higgs boson becomes absolutely stable.  
There emerges the $H$ parity invariance.
Among low energy particles only the Higgs boson is $H$ parity odd,
whereas all other SM particles are $H$ parity even.\cite{HKT, HTU1}

One immediate consequence is that Higgs bosons become the dark 
matter of the universe.  From the WMAP data the Higgs boson mass $m_H$
is estimated around 70$\,$GeV.\cite{HKT}  
This value does not contradict with the LEP2
bound $m_H > 114\,$GeV, as the $ZZH$ coupling exactly vanishes.
In collider experiments Higgs bosons can be produced in pairs.
However, they appear as missing energies and momenta as they do
not decay.\cite{Cheung2010a, Alves2010}

How can we test the model at colliders? We examine this question
by analyzing the precision data of gauge couplings of quarks and leptons,
Higgs pair production at LHC and ILC, KK spectra of various fields,
and production and decay of the first KK modes of gauge bosons at Tevatron and 
LHC.  The model has one free parameter,  the warp factor $z_L$
of the RS space.  It will be found that the present data from colliders
prefer large  $z_L > 10^{15}$ whereas the Higgs mass $\sim 70\,$GeV 
accounting for the dark matter is obtained with $z_L \sim 10^{5}$
in the current model.  The production of the first KK mode of the $Z$ boson
with a mass around 1130$\,$GeV and a width 422$\,$GeV for 
$z_L=10^{15}$ at LHC will be one of the robust signals of the model.

The $SO(5) \times U(1)$ gauge-Higgs unification model at 
$\theta_H = \pm \onehalf \pi$ has similarity to the Higgsless model
in such processes as $WW$ scattering at the tree level as the Higgs boson
contribution is absent due to the vanishing $WWH$ coupling.\cite{Csaki1}
It has been also discussed that the Higgs boson in the model has 
correspondence to the holographic pseudo-Goldstone boson,\cite{Contino1, Gherghetta1} 
resembling the little Higgs model.
The stable Higgs boson serving as dark matter has similarity to
a second Higgs boson in the inert Higgs doublet model with
new parity.\cite{Ma1978}-\cite{Ma2007}
We would like to stress that the current model can make many definitive, 
quantitative predictions by starting from a concrete action, 
to be compared with other predictions.\cite{Agashe3}-\cite{Haba1}

The paper is organized as follows.  The model is introduced in Section 2,
and Kaluza-Klein (KK) expansions of various fields are summarized in Section 3.
In Section 4 gauge couplings of quarks and leptons are determined, and 
are compared with the precision data for forward-backward asymmetries 
in $e^+ e^-$ annihilation on the $Z$ resonance and partial decay widths 
of the $Z$ boson.   In Section 5 pair production of Higgs bosons at
LHC and ILC is examined.  The spectrum of KK towers of gauge bosons,
quarks, and leptons are determined in Section 6.   Couplings of
quarks and leptons to KK gauge bosons are evaluated in Section 7. 
In Section 8 signals of the first KK $Z$ boson at Tevatron and LHC are
discussed.  Section 9 is devoted to conclusions.

\section{Model}
The $SO(5) \times U(1)$ gauge-Higgs unification scenario was  proposed 
by Agashe, Contino, and Pomarol.\cite{ACP}  
We analyze phenomenological consequences of the model given 
in refs.\ \cite{HNU} and \cite{HTU1}.
The model without leptons was introduced in ref.\  \cite{HOOS}.
It has been shown that the model has  $H$ parity invariance which leads to
the stable Higgs boson.\cite{HKT, HTU1}

The model is defined in the Randall-Sundrum  (RS) warped space
with a metric  
\bea
    ds^2 = G_{MN}dx^M dx^N
  = e^{-2\sigma(y)}\eta_{\mu\nu} dx^\mu dx^\nu + dy^2 , \qquad
      \label{metric1}
\eea
where $\eta_{\mu\nu} =\textrm{diag}(-1,1,1,1)$,
$\sigma(y)=\sigma(y+2L) = \sigma(-y)$, and
$\sigma(y)=k|y|$ for $|y|\leq L$.
The Planck and TeV branes are located at $y=0$ and $y=L$, respectively.  
The bulk region $0 < y < L$ is   anti-de Sitter (AdS) spacetime with a 
cosmological constant  $\Lambda = - 6k^2$.  
The warp factor $z_L \equiv e^{kL} \gg 1$  is a  parameter to be specified. 
The Kaluza-Klein (KK) mass scale is given by
$m_\KK = {\pi k}/{(z_L -1)} \sim \pi k z_L^{-1}$, 
which turns out $840 \sim 1470\,$GeV for $z_L= 10^5 \sim 10^{15}$.

The model consists of $SO(5) \times U(1)_X \times SU(3)_c$ gauge fields 
$(A_M, B_M, G_M)$,   bulk fermions $\Psi_a$, brane fermions 
$\hat \chi_{\alpha R}$, and brane scalar $\Phi$.  
The bulk part of the action is given by 
\beqn
&&\hskip -1cm
S_\bulk
= \int d^5x \sqrt{-G} \bigg[ -\tr  {1\over 4} F^{(A)MN} F_{MN}^{(A)}
-  {1\over 4} F^{(B)MN}F_{MN}^{(B)}  \cr
\noalign{\kern 5pt}
&&\hskip 2.8 cm
 - \tr  {1\over 2} F^{(C)MN} F_{MN}^{(C)}
+      \sum_a  i\bar{\Psi}_a {\cal D}(c_a) \Psi_a   \bigg]      ,  \cr
\noalign{\kern 10pt}
&&\hskip -1.cm
{\cal D}(c_a) =   \Gamma^A {e_A}^M
\big( \partial_M +{1\over 8}\omega_{MBC}  [\Gamma^B, \Gamma^C]
  -ig_A A_M \cr
\noalign{\kern 5pt}
&&\hskip 3.cm
-ig_B Q_{Xa} B_M  - i g_C Q^{\rm color} G_M \big)   -c_a \sigma'(y) ~ .
\label{action1}
\eeqn
The gauge fixing and ghost terms associated with the three gauge groups have
been suppressed.  
$F_{MN}^{(A)} = \partial_M A_N -\partial_N A_M -ig_A  [A_M, A_N]$,  
$F_{MN}^{(B)} =\partial_M B_N -\partial_N B_M$, and 
$F_{MN}^{(C)} = \partial_M G_N -\partial_N G_M -ig_C  [G_M, G_N]$. 
$Q^{\rm color} = 1$ or 0   for quark or lepton multiplets, respectively.
The $SO(5)$ gauge fields $A_M$ are decomposed as 
$A_M  
 = \sum_{a_L=1}^3 A^{a_L}_M T^{a_L}+\sum_{a_R =1}^3 A^{a_R}_M T^{a_R}
 +\sum_{\hat{a}=1}^4 A^{\hat{a}}_M T^{\hat{a}}$,
where $T^{a_L, a_R}$ ($a_L, a_R=1,2,3$) and $T^{\hat{a}}$ 
($\hat{a}=1,2,3,4$) are the generators of $SO(4)\simeq SU(2)_L \times SU(2)_R$ 
and $SO(5)/SO(4)$, respectively. 
In the fermion part
$\bar{\Psi} =i\Psi^\dag \Gamma^0$
and Dirac $\Gamma^M$ matrices are given by
\beeq
\Gamma^\mu = 
\begin{pmatrix} & \sigma^\mu \cr \bar{\sigma}^\mu & 
\end{pmatrix}  , \quad
\Gamma^5 =\begin{pmatrix} 1 & \cr & -1  \end{pmatrix}  ,
\quad \sigma^\mu= (1,\vec{\sigma}) , \quad
 \bar{\sigma}^\mu=(-1,\vec{\sigma}) .
\label{matrix1}
\eneq
All of the bulk fermions belong to the vector ({\bf 5}) representation of $SO(5)$. 
The $c_a$ term in Eq.~(\ref{action1}) gives
a bulk kink mass, where
$\sigma'(y)=k\epsilon(y)$ is a periodic
step function with a magnitude $k$.
The dimensionless parameter $c_a$ plays an important  role in controlling profiles
of fermion wave functions.

The orbifold boundary conditions at $y_0=0$ and $y_1=L$  are given by
\beqn
&&\hskip -1.cm
\begin{pmatrix}    A_\mu \cr A_y    \end{pmatrix}  (x,y_j-y)
= P_j  \begin{pmatrix}    A_\mu \cr - A_y  \end{pmatrix}  (x,y_j+y) P_j^{-1} , \cr
\noalign{\kern 10pt}
&&\hskip -1.cm
\begin{pmatrix}    B_\mu \cr B_y    \end{pmatrix}  (x,y_j-y)
= \begin{pmatrix}    B_\mu \cr - B_y  \end{pmatrix}  (x,y_j+y) , \cr
\noalign{\kern 10pt}
&&\hskip -1.cm
\Psi_a (x,y_j-y) = P_j \Gamma^5 \Psi_a (x,y_j+y) , \cr
\noalign{\kern 10pt}
&&\hskip -1.cm
P_j =\textrm{diag} \, (-1,-1,-1,-1,+1)~.
\label{BC1}
\eeqn
The $SO(5)\times U(1)_X$ symmetry is reduced to
$SO(4)\times U(1)_X \simeq SU(2)_L \times SU(2)_R \times U(1)_X$
by the orbifold boundary conditions.
It is known that various orbifold boundary conditions fall into a finite number of 
equivalence classes of boundary conditions.\cite{YH2, HHHK, HHK}  
The physical symmetry of the true vacuum in each equivalence class of boundary
conditions is dynamically determined at the quantum level.

The 4D Higgs field, which is a  doublet both in $SU(2)_L$ and in $SU(2)_R$,
appears as a zero mode in the $SO(5)/SO(4)$ part of the fifth  dimensional component 
of the vector potential $A_y^{\hat a} (x,y)$.  
Without loss of generality one assumes  $\la A_y^{\hat a} \ra  \propto \delta^{a4}$ 
when the EW symmetry is spontaneously broken. 
The zero modes of $A_y^{\hat a}$ ($a=1,2,3$) are absorbed by $W$ and $Z$ bosons.
The Wilson line phase $\theta_H$ is given by 
\beeq
\exp \Big\{\frac{i}{2}\theta_H  \cdot 2\sqrt2 \, T^{\hat{4}}  \Big\}
=\exp \bigg\{ ig_A \int^{L}_{0} dy  \la A_y \ra \bigg\} ~.
\label{Higgs1}
\eneq
The 4D neutral Higgs field $H(x)$ appears as \cite{HK}
\beqn
&&\hskip -1.cm
A^{\hat{4}}_y (x,y)=\big\{ \theta_H f_H + H(x) \big\} u_H(y) +\cdots ~, \cr
\noalign{\kern 10pt}
&&\hskip -1.cm
f_H = \frac{2}{g_A} \sqrt{\frac{k}{z_L^2 -1}}
= \frac{2}{g_w} \sqrt{\frac{k}{L(z_L^2 -1)}} ~.
\label{Higgs2}
\eeqn
Here the wave function of  the 4D Higgs boson is given by
$u_H(y) = [2k/(z_L^2-1)]^{1/2} e^{2ky}$ for $0\le y \le L$ and
$u_H(-y) = u_H(y) = u_H(y + 2L)$.  
$g_w= g_A/\sqrt{L}$ is the dimensionless 4D $SU(2)_L$ coupling.  

For each generation two vector multiplets $\Psi_1$ and $\Psi_2$ for quarks
and two vector multiplets $\Psi_3$ and $\Psi_4$ for leptons are introduced.
Each vector multiplet, $\Psi$, is decomposed into one $(\onehalf, \onehalf)$, 
$\check \Psi$,  and one $(0,0)$ of $SU(2)_L \times SU(2)_R$.  
We denote $\Psi_a$'s , for the third generation, as
\beqn
&&\hskip -1.cm
\Psi_1= \big(  \check \Psi_1 , ~ t'  ~\big)_{2/3} ~,~~
 \check \Psi_1 = \begin{pmatrix} T & t \cr B & b \end{pmatrix} 
 \equiv \bigg( Q_1  ,    q  \bigg) ~,  \cr
\noalign{\kern 10pt}
&&\hskip -1.cm
\Psi_2= \big(  \check \Psi_2 , ~ b'  ~\big)_{-1/3} ~, ~~ 
\check \Psi_2 =  \begin{pmatrix} U &X \cr D & Y \end{pmatrix} 
\equiv \bigg( Q_2  ,  Q_3 \bigg) ~, \cr
\noalign{\kern 10pt}
&&\hskip -1.cm
\Psi_3= \big(  \check \Psi_3 , ~ \tau'  ~\big)_{-1} ~,  ~~
\check \Psi_3 = \begin{pmatrix} \nu_\tau & L_{1X} \cr  \tau  &  L_{1Y} \end{pmatrix}
\equiv \bigg( \ell ,  L_1 \bigg) ~, \cr
\noalign{\kern 10pt}
&&\hskip -1.cm
\Psi_4= \big(  \check \Psi_4 , ~ \nu_\tau '  ~\big)_{0} ~, ~~ 
\check \Psi_4 = \begin{pmatrix}  L_{2X} & L_{3X} \cr L_{2Y}  & L_{3Y} \end{pmatrix}
\equiv \bigg( L_2  ,   L_3 \bigg) ~.
\label{bulkF1}
\eeqn
Subscripts $2/3$ etc.\ represent $U(1)_X$ charges, $Q_X$, of  $\Psi_a$'s.
$q$, $Q_j$, $\ell$, and $L_j$ are $SU(2)_L$ doublets.
The electromagnetic charge $Q_\EM$ is given by
\beeq
Q_\EM = T^{3_L} + T^{3_R} + Q_X ~.
\label{charge1}
\eneq
Each $\Psi_a$ has its bulk mass parameter $c_a$.  Consistent results are
obtained by taking $c_1 = c_2 \equiv c_q$ and $c_3=c_4 \equiv c_\ell$ 
for each generation.

The additional brane fields are introduced on the Planck brane at $y=0$.  
The  brane scalar field $\Phi$ belongs to $(0,{1\over 2})$ of $SU(2)_L \times SU(2)_R$
with $Q_X = - \onehalf$, whereas the right-handed
brane fermions $\hat \chi_{\alpha R}^q$ and $\hat \chi_{\alpha R}^\ell$ belong to
 $({1\over 2}, 0)$.  The brane fermions are
 \beqn
 &&\hskip -1.cm
 \hat \chi_{1R}^q = \begin{pmatrix} \hat T_R \cr \hat B_R \end{pmatrix}_{7/6} , ~~
 \hat \chi_{2R}^q = \begin{pmatrix} \hat U_R \cr \hat D_R \end{pmatrix}_{1/6}, ~~
 \hat \chi_{3R}^q = \begin{pmatrix} \hat X_R \cr \hat Y_R \end{pmatrix}_{-5/6} ,   \cr
 \noalign{\kern 10pt}
&&\hskip -1.cm
 \hat \chi_{1R}^\ell = \begin{pmatrix} \hat L_{1XR} \cr \hat L_{1YR} \end{pmatrix}_{-3/2} , ~~ 
 \hat \chi_{2R}^\ell = \begin{pmatrix} \hat L_{2XR} \cr \hat L_{2YR} \end{pmatrix}_{1/2} , ~~ 
 \hat \chi_{3R}^\ell = \begin{pmatrix} \hat L_{3XR} \cr \hat L_{3YR} \end{pmatrix}_{-1/2} .
 \label{braneF1}
 \eeqn
Subscripts $7/6$ etc. represent $Q_X$ charges of  $\hat \chi_R$'s.
The brane part of the action is given by
\beqn
&&\hskip -1.cm
S_\brane
= \int d^5 x  \sqrt{-G} ~   \delta(y) \bigg\{
   -(D_\mu \Phi)^\dag D^\mu \Phi  -\lambda_\Phi  (\Phi^\dag \Phi -w^2)^2  \cr
\noalign{\kern 5pt}
&&\hskip -.5cm
+ \sum_{\alpha=1}^3 \Big( \hat{\chi}_{\alpha R}^{q \dag}
  \,  i \bar{\sigma}^\mu   D_\mu \hat{\chi}_{\alpha R}^q
   +  \hat{\chi}_{\alpha R}^{\ell \dag}
\,  i \bar{\sigma}^\mu   D_\mu \hat{\chi}_{\alpha R}^\ell  \Big)     \cr
\noalign{\kern 5pt}
&&\hskip -.5cm
-   i\Big[ \kappa_1^q \, \hat{\chi}_{1 R}^{q \dag} \check \Psi_{1L} \tilde \Phi 
  + \tilde \kappa^q \,  \hat{\chi}_{2 R}^{q \dag} \check \Psi_{1L}  \Phi
  + \kappa_2^q \,  \hat{\chi}_{2 R}^{q \dag} \check \Psi_{2L} \tilde \Phi 
  + \kappa_3^q \,  \hat{\chi}_{3 R}^{q \dag} \check \Psi_{2L} \Phi 
  - (\hbox{h.c.}) \Big] \cr
\noalign{\kern 10pt}
&&\hskip -.5cm
-   i\Big[ \tilde \kappa^\ell \, \hat{\chi}_{3 R}^{\ell \dag} \check \Psi_{3L} \tilde \Phi 
  + \kappa_1^\ell \,  \hat{\chi}_{1 R}^{\ell \dag} \check \Psi_{3L}  \Phi
  + \kappa_2^\ell \,  \hat{\chi}_{2 R}^{\ell \dag} \check \Psi_{4L} \tilde \Phi 
  + \kappa_3^\ell \,  \hat{\chi}_{3 R}^{\ell \dag} \check \Psi_{4L} \Phi 
  - (\hbox{h.c.}) \Big] \bigg\} ~, \cr
\noalign{\kern 5pt}
&&\hskip 0.5cm
D_\mu \Phi = \Big( \dd_\mu - i   g_A \sum_{a_R=1}^3  A_\mu^{a_R} T^{a_R}
    + i  {1\over 2}g_B B_\mu \Big) \Phi  ~, 
     ~~ \tilde \Phi = i \sigma_2 \Phi^* ~,  \cr
\noalign{\kern 0pt}
&&\hskip 0.5cm
D_\mu \hat \chi = \Big( \dd_\mu  - i  g_A \sum_{a_L=1}^3  A_\mu^{a_L} T^{a_L}
      -i  Q_X g_B B_\mu  - i g_C Q^{\rm color} G_\mu \Big) \hat \chi  ~.
\label{action2}
\eeqn
The  action $S_\brane$ is manifestly invariant under 
$SU(2)_L \times SU(2)_R \times U(1)_X$.  
The Yukawa couplings above exhaust all possible ones preserving the symmetry.

The non-vanishing vev $\la \Phi^t \ra = (0, w)$ has two important consequences.  
It is assumed only that $w \gg m_\KK$.  
Firstly the $SU(2)_R \times U(1)_X$ symmetry is spontaneously broken down 
to $U(1)_Y$  and the zero modes of four-dimensional gauge fields of 
$SU(2)_R \times U(1)_X$ become massive except for the $U(1)_Y$ part.  
They acquire masses of $O(m_\KK)$ as a result of the effective change of 
boundary conditions for low-lying modes in the Kaluza-Klein towers.
Secondly the non-vanishing vev $w$ induces mass couplings between
brane fermions and bulk fermions;
\beqn
&&\hskip -1.cm
S_\brane^{\rm mass} = \int d^5 x  \sqrt{-G} ~   \delta(y) \bigg\{
-\sum_{\alpha=1}^3  
     i \mu_\alpha^q  (\hat{\chi}_{\alpha R}^{q\dag} Q_{\alpha L}
    -Q_{\alpha L}^\dag \hat{\chi}_{\alpha R}^q) 
    -i \tilde{\mu}^q (\hat{\chi}_{2R}^{q \dag} q_L   - q_L^\dag \hat{\chi}_{2R}^q) \cr
\noalign{\kern 5pt}
&&\hskip 3.cm
- \sum_{\alpha=1}^3  
    i \mu_\alpha^\ell  (\hat{\chi}_{\alpha R}^{\ell \dag} L_{\alpha L}
    -L_{\alpha L}^\dag \hat{\chi}_{\alpha R}^\ell ) 
    -i \tilde{\mu}^\ell  (\hat{\chi}_{3R}^{\ell \dag} \ell_L   
    -\ell_L^\dag \hat{\chi}_{3R}^\ell)     \bigg\} ~,  \cr
\noalign{\kern 10pt}
&&\hskip 2.cm
\frac{\mu_\alpha^q}{\kappa_\alpha^q} = \frac{\tilde \mu^q}{\tilde \kappa^q}
= \frac{\mu_\alpha^\ell}{\kappa_\alpha^\ell} = \frac{\tilde \mu^\ell}{\tilde \kappa^\ell}
=  w ~,
\label{action3}
\eeqn
Assuming that all $\mu^2 \gg m_\KK$,  all of the exotic zero modes of
the bulk fermions acquire large masses of $O(m_\KK)$.  
It has been shown that all of the 4D anomalies associated with 
$SU(2)_L \times SU(2)_R \times U(1)_X$ gauge symmetry are cancelled.\cite{HNU}
The $SU(2)_L \times U(1)_Y$  is further broken down to $U(1)_\EM$ 
by the Hosotani mechanism.
The spectrum of the resultant light particles are the same as  in the standard model.
The generation mixing can be explained by considering matrix couplings of
$\kappa_j$ and $\tilde \kappa$, particularly of 
$\kappa_2^q, \tilde \kappa^q,  \kappa_3^l, \tilde \kappa^l$.

The parameters of the model relevant for low energy physics are 
$k$, $z_L=e^{kL}$, $g_A$, $g_B$,  the bulk mass parameters 
$(c_q, c_\ell)$ and the brane mass ratios 
$(\tilde \mu^q / \mu^q_2, \tilde \mu^\ell / \mu^\ell_3)$.  
All other parameters are irrelevant at low energies, provided that
$w$,  $\mu^2$'s are much larger than $m_\KK$.
The value of $\theta_H$ is determined dynamically to be
$\pm \onehalf \pi$ where the EW symmetry is spontaneously broken.\cite{HTU1} 
Three of the four parameters $k$, $z_L=e^{kL}$, $g_A$, $g_B$ are determined
from the $Z$ boson mass $m_Z$, the weak gauge coupling $g_w$, and 
the Weinberg angle $\sin^2 \theta_W$.  The one parameter, say, $z_L$ remains free.

When the generation mixing is turned off in the fermion sector, 
the bulk mass $c_q$ and the ratio $\tilde \mu^q / \mu^q_2$ in each generation
are determined from the two quark masses, and 
$c_\ell$ and $ \tilde \mu^\ell / \mu^\ell_3$ from the two lepton masses.
As $m_{\nu_e} \ll m_e$,  all of the results 
discussed below do not depend on the unknown value of $m_{\nu_e}$ very much.
The generation mixing can be incorporated by considering 3-by-3 matrices
for the brane couplings $\kappa$'s, or equivalently 
for the brane masses $\mu$'s.

Once the value of $z_L$ is specified, all the relevant parameters of the model 
are determined.  The spectra of particles and their KK towers, 
their wave functions in the fifth dimension, and all interaction couplings can 
be predicted.    The mass of the 4D Higgs boson, $m_H$, is determined from 
the effective potential $V_\eff (\theta_H)$.
It was found that $m_H$ is about $70 \sim 135\,$GeV for
$z_L = 10^{5} \sim 10^{15}$.\cite{HTU1}
Conversely the remaining one parameter
$z_L$ is fixed, once the Higgs boson mass $m_H$ is given.

As typical reference values we take  the warp factors $z_L = 10^5, 10^{10}, 10^{15}$.
The values in Table~\ref{tab:input} are taken as input parameters. 
The masses of quarks and charged leptons except for 
$t$ quark are quoted from Ref.~\cite{Xing:2007fb}.
The masses of $Z$ boson and $t$ quark are
the central values in the Particle Data Group review~\cite{Amsler:2008zzb}.
The couplings $\alpha$ and $\alpha_s$ are 
also quoted from Ref.~\cite{Amsler:2008zzb}.
In the present analysis, the neutrino masses have negligible effects.

The parameter $\sin^2 \theta_W$ is determined from 
$\chi^2$ fit of
forward-backward asymmetries  in $e^+ e^-$ annihilation and branching ratios in the
$Z$ decay as explained below.   We find the best fit with
$\sin^2 \theta_W =0.2284,  0.2303, 0.2309$ for 
$z_L = 10^{5}, 10^{10}, 10^{15}$, respectively.   
Since complete one-loop analysis is not available in the 
gauge-Higgs unification scenario at the moment,  there remains
ambiguity in the value of $\sin^2 \theta_W$.

\begin{table}[htb]
\begin{center}
\caption{Input parameters for the masses and couplings of the model.
The masses are  in an unit of GeV.   All masses except for $m_t$ are
at the $m_Z$ scale.}
\label{tab:input}
\vskip 7pt
\begin{tabular}{|c|c|}
\hline 
$m_Z$ & 91.1876 
\\
\hline
$m_u$ &$1.27 \times 10^{-3}$ 
\\
\hline
$m_c$ &0.619 
\\
\hline
$m_t$ &171.17 
\\
\hline
$m_d$ & $2.90 \times 10^{-3}$ 
\\
\hline
$m_s$ & 0.055 
\\
\hline
$m_b$ & 2.89 
\\
\hline
\end{tabular}
\hskip 5pt
\begin{tabular}{|c|c|}
\hline 
$\alpha_s (m_Z)$ & 0.1176 
\\
\hline
$m_{\nu_e}$ &$1 \times 10^{-12}$
\\
\hline
$m_{\nu_\mu}$ &$9 \times 10^{-12}$
\\
\hline
$m_{\nu_\tau}$ &$5.0309 \times 10^{-11}$
\\
\hline
$m_e$ &$0.486570161 \times 10^{-3}$
\\
\hline
$m_\mu$ &$102.7181359 \times 10^{-3}$
\\
\hline
$m_\tau$ &1.74624
\\
\hline
\end{tabular}
\end{center}
\end{table}

\section{Kaluza-Klein expansion}

With the orbifold boundary condition (\ref{BC1}) the effective potential 
$V_\eff (\theta_H)$ is minimized at $\theta_H = \onehalf \pi$.  
To develop perturbation theory around $\theta_H = \onehalf \pi$,
it is most convenient to move to the twisted gauge 
$\tilde A_M = \Omega A_M \Omega^{-1}  + (i/g_A) \Omega \dd_M \Omega^{-1}$
in which $\la \tilde A_y \ra = 0$, or $\tilde \theta_H = 0$.  We choose $\Omega$
preserving the boundary condition at the TeV brane;
\beeq
\Omega(y) = \exp \bigg\{ \frac{i \pi g_A f_H}{2} \int_y^L dy' \, u_H(y') \cdot 
T^{\hat 4} \bigg\} ~.
\label{gauge1}
\eneq
In the twisted gauge the orbifold boundary condition $\{ P_0, P_1 \}$ is
changed to $\{ \tilde P_0,  \tilde P_1 \}$ where 
\beqn
&&\hskip -1cm
\tilde P_0 = \Omega(-y) P_0 \Omega(y)^{-1} = \diag (-1,-1,-1,+1, -1)
\not= P_0  ~, \cr
\noalign{\kern 5pt}
&&\hskip -1cm
\tilde P_1 = \Omega(L - y) P_1 \Omega(L + y)^{-1} = \diag (-1,-1,-1,-1, +1)
= P_1 ~.
\label{BC2}
\eeqn
The two sets $\{ P_0, P_1 \}$ and $\{ \tilde P_0,  \tilde P_1 \}$ are in the same 
equivalence class of boundary conditions.\cite{YH2, HHHK, HHK, Hosotani2003}   
Although the boundary conditions are different, physics remains
the same as a result of dynamics of the Wilson line phase.
Note that $\Omega(L) = 1$, but
\beeq
\Omega(0) = \begin{pmatrix}
1 \cr &1 \cr && 1 \cr &&& 0 & 1 \cr &&& -1 & 0 \end{pmatrix} 
\label{gauge2}
\eneq
so that the brane interactions take more complicated form than in the original
gauge.

In the previous paper it was shown that the model has 
$H$ parity ($P_H$) invariance, and
$H$ parity is assigned to all 4D fields.\cite{HTU1}
$P_H$ interchanges $SU(2)_L$ and $SU(2)_R$ and flips the 
sign of $T^{\hat 4}$ in the bulk.  
$P_H$ transformation is generated by 
$T^\alpha \go \Omega_H T^\alpha \Omega_H^{-1}$ where
$\Omega_H = \diag (1,1,1,-1,1)$ in the twisted gauge.
The $P_H$ symmetry is similar to the  $P_{LR}$ symmetry
discussed by Agashe, Contino, Da Rold and Pomarol  \cite{Agashe2006}, 
which  protects the $T$ parameter and 
$Z b \bar b$ coupling from radiative corrections.
The neutral Higgs boson is the lightest particle of odd $P_H$ so that
it becomes stable.

In the twisted gauge the four-dimensional components of gauge fields are 
expanded as
\beqn
&&\hskip -1.cm
\tilde{A}_\mu(x,z) = \hat W_\mu + \hat W_\mu^\dagger + \hat Z_\mu +
\hat  A^{\gamma}_\mu +  \hat W'_\mu + \hat W_\mu^{\prime\dagger}
+ \hat Z'_\mu +\hat A^{\hat 4}_\mu ~ ,  
\cr
\noalign{\kern 10pt}
&&\hskip 0.cm
\hat W_\mu = \sum_n  W_\mu^{(n)} \Big\{ h_{W^{(n)}}^L T^{-_L} 
+ h_{W^{(n)}}^R T^{-_R} + h_{W^{(n)}}^\wedge T^{\hat{-}} \Big\} ~, \cr
 \noalign{\kern 5pt}
&&\hskip 0.cm
\hat Z_\mu =  \sum_n  Z^{(n)}_\mu \Big\{ h_{Z^{(n)}}^L T^{3_L} 
+h_{Z^{(n)}}^R T^{3_R} + h_{Z^{(n)}}^\wedge T^{\hat{3}} \Big\}~,  \cr
 \noalign{\kern 5pt}
&&\hskip 0.cm
\hat  A^{\gamma}_\mu = 
\sum_n  A^{\gamma (n)}_\mu \Big\{  h_{\gamma^{(n)}}^L T^{3_L}
 + h_{\gamma^{(n)}}^R T^{3_R} \Big\}  ~, \cr
 \noalign{\kern 10pt}
&&\hskip 0.cm
\hat W'_\mu = \sum_n  W_\mu^{\prime (n)} \Big\{ h_{W^{\prime (n)}}^L T^{-_L} 
+ h_{W^{\prime (n)}}^R T^{-_R}  \Big\} ~, \cr
 \noalign{\kern 5pt}
&&\hskip 0.cm
\hat Z'_\mu =  \sum_n  Z^{\prime (n)}_\mu \Big\{ h_{Z^{\prime (n)}}^L T^{3_L} 
+h_{Z^{\prime (n)}}^R T^{3_R}  \Big\}~,  \cr
\noalign{\kern 5pt}
&&\hskip 0.cm
\hat  A^{\hat 4}_\mu = 
\sum_n  A^{\hat 4 (n)}_\mu h_{A^{(n)}} T^{\hat 4}~, \cr
\noalign{\kern 10pt}
&&\hskip -1.cm
\tilde{B}_\mu(x,z) = 
\sum_n  Z^{(n)}_\mu h_{Z^{(n)}}^B 
+ \sum_n  A_\mu^{\gamma (n)} h_{\gamma^{(n)}}^B   ~.
\label{expansion1}
\eeqn
Here $T^{\pm} =(T^1 \pm i T^2)/\sqrt{2}$.
The $W$ and $Z$ bosons and the photon $\gamma$ correspond to
$W_\mu^{(0)}$, $Z_\mu^{(0)}$ and $A_\mu^{\gamma (0)}$, respectively.  
Unless confusion arises, we will omit the superscript $(0)$ for representing
the lowest mode.  
The mixing angle between $SO(5)$ and $U(1)_X$ is related to the Weinberg angle
by $\sin^2 \theta_W \equiv s_\phi^2 /(1+s_\phi^2)$ 
where $s_\phi =g_B /\sqrt{g_A^2 + g_B^2}$.
All mode functions $h(z)$ are tabulated in Appendix A.
They are expressed in terms of Bessel functions
\bea
&&\hskip -1cm 
C(z;\lambda) =
      {\pi\over 2}\lambda z z_L F_{1,0}(\lambda z, \lambda z_L) ~, \quad
C'(z;\lambda) =
      {\pi\over 2}\lambda^2 z z_L F_{0,0}(\lambda z, \lambda z_L) ~ , \cr
\noalign{\kern 5pt}
&&\hskip -1cm 
S(z;\lambda) =
      -{\pi\over 2}\lambda z  F_{1,1}(\lambda z, \lambda z_L) ~ , \quad
S'(z;\lambda) =
      -{\pi\over 2}\lambda^2 z F_{0,1}(\lambda z,  \lambda z_L) ~,   \cr
\noalign{\kern 5pt}
&&\hskip -1cm 
F_{\alpha,\beta}(u,v) =
    J_\alpha (u) Y_\beta(v)   -Y_\alpha(u) J_\beta(v) ~.
\label{BesselF1}
\eea
%$N, h_\gamma  \propto C(z; \lambda)$ and $D, h_A  \propto S(z; \lambda)$.
For the photon ($\lambda_0 = 0$), $h_{\gamma^{(0)}}^L=h_{\gamma^{(0)}}^R$ is constant.

The mass spectrum $m_n = k \lambda_n$ of each KK tower is determined by
the corresponding eigenvalue equations:
\beqn
W_\mu^{(n)} &:& 2 S(1; \lambda_n) C' (1; \lambda_n) + \lambda_n  =0 ~, \cr
\noalign{\kern 5pt}
Z_\mu^{(n)} &:& 
2 S(1; \lambda_n) C' (1; \lambda_n) + \lambda_n  (1+ s_\phi^2)   =0 ~, \cr
\noalign{\kern 5pt}
W_\mu^{\prime (n)}, Z_\mu^{\prime (n)} &:& C(1; \lambda_n) = 0 ~, \cr
\noalign{\kern 5pt}
A_\mu^{\gamma(n)} &:& C' (1; \lambda_n) = 0 ~, \cr
\noalign{\kern 5pt}
A_\mu^{\hat{4}(n)}  &:& S(1; \lambda_n) = 0 ~.
\label{spectrum1}
\eeqn
The Weinberg angle $\theta_W$ is determined by global fit of various quantities.  
In the present paper $\sin^2 \theta_W$ is determined from $\chi^2$ fit of
forward-backward asymmetries in $e^+ e^-$ annihilation and $Z$ boson decay.
With $m_Z$ and $z_L$ as an input, the AdS curvature
$k$ and the $W$ boson mass at the tree level, $m_W^{\rm tree}$,  are determined.

Similarly the fifth-dimensional components $A_z$ and $B_z$ are expanded as
\beqn
&&\hskip -1.cm
\tilde{A}_z (x,z)    =
\sum_{a=1}^3    \sum_{n=1}^\infty 
 S^{a (n)} h_{S}^{LR}(\lambda_n)  \frac{T^{a_L} + T^{a_R}}{\sqrt{2}}
+ \sum_{n=0}^\infty  H^{(n)} h_{H}^\wedge (\lambda_n)  T^{\hat{4}}    \cr
\noalign{\kern 5pt}
&&\hskip .5cm    + \sum_{a=1}^3  \sum_{n=1}^\infty   
D_-^{a(n)}   h_{D}^{LR}(\lambda_n)  \frac{T^{a_L} -T^{a_R}}{\sqrt{2}}
+ \sum_{a=1}^3  \sum_{n=1}^\infty  
\hat D^{a (n)}  h_{D}^\wedge (\lambda_n)  T^{\hat{a}}~,  \cr
\noalign{\kern 10pt}
&&\hskip -1.cm
\tilde{B}_z (x,z)  = \sum_{n=1}^\infty    B^{(n)} h_{B} (\lambda_n) ~.
\label{expansion2}
\eeqn
% Here $h_{S}^{LR}, h_{D}^{LR}, h_B \propto C'(z; \lambda)$ and
% $h_{H}^\wedge, h_{D}^\wedge  \propto S'(z; \lambda)$.
$H(x) = H^{(0)} (x)$ is the 4D neutral Higgs boson.
The mass spectrum of each KK tower is given by
\beqn
S^{a (n)}, B^{(n)} &:& C'(1; \lambda_n) = 0 ~, \cr
\noalign{\kern 5pt}
D_-^{a(n)} &:& C(1; \lambda_n) =0 ~, \cr
\noalign{\kern 5pt}
\hat D^{a (n)} &:& S'(1; \lambda_n) =0 ~, \cr
\noalign{\kern 5pt}
H^{(n)} &:& S(1; \lambda_n) = 0 ~, 
\label{spectrum2}
\eeqn

For the bulk fields $H$ parity is assigned from the behavior under
the transformation 
$\{ T^{a_L}, T^{a_R}, T^{\hat a}, T^{\hat 4} \}
\go \{ T^{a_R}, T^{a_L}, T^{\hat a}, -T^{\hat 4} \} $.
It  interchanges $SU(2)_L$ and $SU(2)_R$ and flips the direction of
$T^{\hat 4}$.  Accordingly $P_H$ odd fields are
\beeq
P_H ~ \hbox{odd} ~:~
W^{\prime (n)}_\mu , ~ Z^{\prime (n)}_\mu, 
~ A_\mu^{\hat 4 (n)},  ~ H^{(n)},  ~ D_-^{a(n)} .
\label{Hodd1}
\eneq
Other fields are $P_H$ even.  

As for the fermions, a consistent model is obtained with the bulk mass 
parameters $c_1=c_2 \equiv c_q$ and $c_3=c_4 \equiv c_\ell$.
Let us first consider the  multiplets
containing quarks, namely, $\Psi_1$ and $\Psi_2$ in (\ref{bulkF1}) and
${\hat\chi}^q_{1R}$, ${\hat\chi}^q_{2R}$,  ${\hat\chi}^q_{3R}$,   
in (\ref{braneF1}).  They are classified in terms of electric charge 
$Q_E=\frac{5}{3}$, $\frac{2}{3}$, $-\frac{1}{3}$, $-\frac{4}{3}$.  

We recall that components of $\check \Psi$ in (\ref{bulkF1}) are related to
the components $\Psi^k$ ($k= 1 \sim 5$) in the vectorial representation by
\beeq
\check{\Psi} = 
\begin{pmatrix}
      \check{\Psi}_{11} & \check{\Psi}_{12} \cr
      \check{\Psi}_{21} & \check{\Psi}_{22} 
\end{pmatrix} 
= - \frac{1}{\sqrt{2}} 
\begin{pmatrix}
     \Psi^2 +i\Psi^1 & -\Psi^4 - i\Psi^3 \\
     \Psi^4 -i\Psi^3 &  \Psi^2 -i\Psi^1 \\
\end{pmatrix} ~.
\label{bulkF2}
\eneq
$\Psi^4$ and $\Psi^5$ couple with $A_z^{\hat 4}$ or $\theta_H$.
Conversely we have, for the third generation  in the twisted gauge, 
\beeq
\begin{pmatrix} \tilde \Psi^1_1 \cr \tilde \Psi^2_1 \cr \tilde \Psi^3_1 \cr 
\tilde \Psi^4_1 \cr  \tilde \Psi^5_1   \end{pmatrix} =
\begin{pmatrix} i( \tilde T - \tilde b)/\sqrt{2} \cr -( \tilde T + \tilde b)/\sqrt{2} \cr
-i (\tilde B + \tilde t)/\sqrt{2} \cr - (\tilde B - \tilde t)/\sqrt{2} \cr t' \end{pmatrix}  , ~~
\begin{pmatrix} \tilde \Psi^1_2 \cr \tilde \Psi^2_2 \cr \tilde \Psi^3_2 \cr 
\tilde \Psi^4_2 \cr  \tilde \Psi^5_2   \end{pmatrix} =
\begin{pmatrix} i( \tilde U - \tilde Y)/\sqrt{2} \cr -( \tilde U + \tilde Y)/\sqrt{2} \cr
-i (\tilde D + \tilde X)/\sqrt{2} \cr - (\tilde D - \tilde X)/\sqrt{2} \cr b' \end{pmatrix}  , ~~
\label{bulkF3}
\eneq
$\Omega_H$ transformation gives 
 $(\tilde \Psi^1,  \tilde \Psi^2,\tilde \Psi^3,\tilde \Psi^4,\tilde \Psi^5)
\go (\tilde \Psi^1,  \tilde \Psi^2,\tilde \Psi^3, -\tilde \Psi^4,\tilde \Psi^5)$.
The $\tilde \Psi^4$ component is $P_H$ odd, whereas other components
are $P_H$ even.  

The $Q_E = \frac{5}{3}$ sector consists of $T$ in $\Psi_1$ and 
$\hat T_R$ in $\hat \chi^q_{1R}$.  
The $Q_E = - \frac{4}{3}$ sector consists of $Y$ in $\Psi_2$ and 
$\hat Y_R$ in $\hat \chi^q_{3R}$.
There are no light modes in these two sectors.

The $Q_E = \frac{2}{3}$ sector consists of $B$, $t$, $t'$ in $\Psi_1$,
$U$  in $\Psi_2$,  $\hat B_R$ in $\hat \chi^q_{1R}$ and
$\hat U_R$ in $\hat \chi^q_{2R}$.  
The bulk fermions have the following  Kaluza-Klein expansion.
\beqn
\begin{pmatrix}  \tilde{U}_L \\  (\tilde{B}_L + \tilde{t}_L)/\sqrt{2} \\
     \tilde{t}_L'    \end{pmatrix}  (x,z)
&\myeq &  \sqrt{k}  \sum_{n=0}^\infty 
\begin{pmatrix}   a_U^{(n)}  C_L(z;\lambda_n , c_t) \\
   a_{B+ t}^{(n)} C_L(z; \lambda_n , c_t) \\
   a_{t'}^{(n)}  S_L(z;\lambda_n, c_t) \end{pmatrix}  \psi^{(n)}_{\frac{2}{3} (+), L}(x) ~ , \cr
\noalign{\kern 5pt}
\begin{pmatrix}  \tilde{U}_R \\  (\tilde{B}_R + \tilde{t}_R)/\sqrt{2} \\
     \tilde{t}_R'    \end{pmatrix}  (x,z)
&\myeq &   \sqrt{k}  \sum_{n=0}^\infty 
\begin{pmatrix}   a_U^{(n)}  S_R(z;\lambda_n , c_t) \\
   a_{B+ t}^{(n)} S_R(z; \lambda_n , c_t) \\
   a_{t'}^{(n)}  C_R(z;\lambda_n, c_t) \end{pmatrix}  \psi^{(n)}_{\frac{2}{3} (+), R}(x) ~, \cr
\noalign{\kern 5pt}
\begin{pmatrix}
 (\tilde{B}_L - \tilde{t}_L) /\sqrt{2} \cr
 (\tilde{B}_R - \tilde{t}_R) /\sqrt{2}
 \end{pmatrix}  (x,z)   &\myeq &
 \sqrt{k}  \sum_{n=1}^\infty  a_{B - t}^{(n)} 
 \begin{pmatrix}
 C_L(z; \lambda_n, c_t ) \,  t^{(n)}_{(-), L}(x) \cr
 S_R(z; \lambda_n, c_t ) \,  t^{(n)}_{(-), R}(x)  \end{pmatrix}   ~ .
\label{FermionKK1}
\eeqn
Here $c_t$ is the bulk kink mass for the third generation $(\Psi_1, \Psi_2)$, and 
\beqn
&&\hskip -1cm
\begin{pmatrix} C_L \cr S_L \end{pmatrix}  (z;\lambda, c)
= \pm \frac{\pi}{2} \lambda\sqrt{zz_L}
   F_{c+{1\over 2},c\mp{1\over 2}}  (\lambda z, \lambda z_L) ~, \cr
\noalign{\kern 10pt}
&&\hskip -1cm  
\begin{pmatrix} C_R \cr S_R \end{pmatrix}  (z;\lambda, c)
= \mp \frac{\pi}{2} \lambda\sqrt{zz_L}
F_{c-{1\over 2},c\pm {1\over 2}} 
 (\lambda z, \lambda z_L)~.
\label{Bessel2}
\eeqn
$\psi^{(n)}_{\frac{2}{3} (+)}(x)$ fields are $P_H$ even, while $t^{(n)}_{(-)}(x)$
fields are $P_H$ odd.  $\{ \psi^{(n)}_{\frac{2}{3} (+)}(x) \}$ contains
three KK towers, including the KK tower $t^{(n)}_{(+)}(x)$ of the top quark.
The brane fields $\hat B_R$ and $\hat U_R$ can be expressed in terms of 
the bulk fields.  

The spectrum $\lambda_n$ and mode coefficients $a^{(n)}$ of the $P_H$-even 
towers satisfy
\beeq
\det \hat K = 0 ~~,~~
\hat K ~ \begin{pmatrix}
a_U^{(n)} \cr  \myfrac{1}{2}~ a_{B + t}^{(n)}\cr   
\myfrac{1}{\sqrt{2}} ~ a_{t'}^{(n)} 
\end{pmatrix} = 0 ~, 
\label{FermionKK2}
\eneq
where
\beqn
&& \hskip -1.cm
\hat K = 
\begin{pmatrix}
\lambda_n S_R- \myfrac{\mu_2^2}{2k} C_L
& - \myfrac{\mu_2 \tilde\mu}{2k} C_L 
& \myfrac{\mu_2 \tilde\mu}{2k} S_L  \cr
0 &  \lambda_n S_R - \myfrac{\mu_1^2}{2k} C_L
& \big( \lambda_n C_R - \myfrac{\mu_1^2}{2k} S_L \big)  \cr
- \myfrac{\mu_2 \tilde\mu}{2k} C_L
& \lambda_n S_R - \myfrac{\tilde\mu^2}{2k} C_L 
&-  \big( \lambda_n C_R - \myfrac{\tilde\mu^2}{2k} S_L \big)
\end{pmatrix} 
 ~, \cr
\noalign{\kern 10pt}
&& \hskip -1.cm
C_{L,R} = C_{L,R}(1; \lambda_n, c_t) ~~,~~
S_{L,R} = S_{L,R} (1; \lambda_n, c_t) ~~.
\label{FermionKK3}
\eeqn
Here we have suppressed a superscript $q$ in $\mu_j^q$. 
There is one light mode (the top quark) with $ m_t = k \lambda_{t,0} \ll m_\KK$.  
When $\mu_j^2, \tilde \mu^2 \gg m_\KK$, the spectrum of 
the top quark tower satisfies
\beeq
2 \bigg(1 + \frac{\tilde \mu^2}{\mu_2^2} \bigg)  
S_R (1; \lambda_{t,n},  c_t) S_L (1; \lambda_{t,n},  c_t) + 1 =0
\label{FermionKK4}
\eneq
for $k \lambda_{t,n} \ll m_\KK$.
A similar relation is obtained for the bottom quark  $m_b = k \lambda_{b,0}$;
\beeq
2 \bigg( 1 + \frac{\mu_2^2}{\tilde \mu^2} \bigg)  
S_R (1; \lambda_{b,n},  c_t) S_L (1; \lambda_{b,n},  c_t) + 1 = 0
\label{FermionKK5}
\eneq
for $k \lambda_{b,n} \ll m_\KK$.
With $(m_t, m_b)$ given, Eqs. (\ref{FermionKK4}) and (\ref{FermionKK5})
determine the bulk mass $c_t$ and the ratio $\tilde \mu^2/\mu_2^2$.  
We note that $\tilde \mu^2/\mu_2^2 \sim m_b/m_t$ for $m_b \ll m_t$.
The spectrum of the KK tower $t^{(n)}_{(-)}(x)$ is determined by 
$C_L(1; \lambda_n, c_t) =0$.

Parallel arguments apply to the $Q_E = - \frac{1}{3}$ sector, which consists of 
$b$ in $\Psi_1$, $D, X, b'$ in $\Psi_2$, $\hat D_R$ in $\hat \chi^q_{2R}$ and
$\hat X_R$ in $\hat \chi^q_{3R}$.  The bulk fields are expanded as
\beqn
\begin{pmatrix}  \tilde{b}_L \\  (\tilde{D}_L + \tilde{X}_L)/\sqrt{2} \\
     \tilde{b}_L'    \end{pmatrix}  (x,z)
&\myeq &  \sqrt{k}  \sum_{n=0}^\infty 
\begin{pmatrix}   a_b^{(n)}  C_L(z;\lambda_n , c_t) \\
   a_{D+X}^{(n)} C_L(z; \lambda_n , c_t) \\
   a_{b'}^{(n)}  S_L(z;\lambda_n, c_t) \end{pmatrix}  \psi^{(n)}_{-\frac{1}{3} (+), L}(x) ~ , \cr
\noalign{\kern 5pt}
\begin{pmatrix}  \tilde{b}_R \\  (\tilde{D}_R + \tilde{X}_R)/\sqrt{2} \\
     \tilde{b}_R'    \end{pmatrix}  (x,z)
&\myeq &   \sqrt{k}  \sum_{n=0}^\infty 
\begin{pmatrix}   a_b^{(n)}  S_R(z;\lambda_n , c_t) \\
   a_{D+X}^{(n)} S_R(z; \lambda_n , c_t) \\
   a_{b'}^{(n)}  C_R(z;\lambda_n, c_t) \end{pmatrix}  \psi^{(n)}_{-\frac{1}{3} (+), R}(x) ~, \cr
\noalign{\kern 5pt}
\begin{pmatrix}
 (\tilde{D}_L - \tilde{X}_L) /\sqrt{2} \cr
 (\tilde{D}_R - \tilde{X}_R) /\sqrt{2}
 \end{pmatrix}  (x,z)   &\myeq &
 \sqrt{k}  \sum_{n=1}^\infty  a_{D-X}^{(n)} 
 \begin{pmatrix}
 C_L(z; \lambda_n, c_t ) \,  b^{(n)}_{(-), L}(x) \cr
 S_R(z; \lambda_n, c_t ) \,  b^{(n)}_{(-), R}(x)  \end{pmatrix}   ~ .
\label{FermionKK6}
\eeqn
The equations and relations in the  $Q_E = - \frac{1}{3}$ sector are obtained
from those in the  $Q_E = \frac{2}{3}$ sector by replacing
$(U, B, t, t')$ and $(\mu_1, \mu_2, \tilde \mu)$
by $(b, D, X, b')$ and  $(\mu_3, \tilde\mu, \mu_2)$, respectively.

Similar relations are obtained in the lepton sector.  
The generation mixing is incorporated by considering $\mu_2, \tilde \mu$
in matrices.

\section{4D gauge couplings \label{sec:4dcoupling}}

The 4D gauge couplings are obtained by performing
overlapping integrals of wave functions.
Generalizing the argument in Ref.~\cite{HNU}, one can write, 
for the $t$ and $b$ quarks and the $\tau$ and $\nu_\tau$ leptons
in the third generation, 
the couplings of the photon,  $W$ boson, $Z$ boson, and gluon towers as
\beqn
&&\hskip -1.cm
\sum_n A_\mu^{\gamma (n)} \Big\{
\twothird  \big(g_{t L}^{\gamma^{(n)}} \bar{t}_L \gamma^\mu t_L
%\frac{2}{3} \,  \big(g_{t L}^{(\gamma_n)} \bar{t}_L \gamma^\mu t_L
      +g_{t R}^{\gamma^{(n)}} \bar{t}_R \gamma^\mu t_R \big)
-\onethird  \big(g_{b L}^{\gamma^{(n)}} \bar{b}_L \gamma^\mu b_L
    + g_{b_R}^{\gamma^{(n)}} \bar{b}_R \gamma^\mu b_R \big)   \cr
\noalign{\kern 10pt}
&&\hskip 1.cm
-   \big(g_{\tau L}^{\gamma^{(n)}} \bar{\tau}_L \gamma^\mu \tau_L
    + g_{\tau_R}^{\gamma^{(n)}} \bar{\tau}_R \gamma^\mu \tau_R \big) \Big\}  \cr
\noalign{\kern 10pt}
&&\hskip -1.4cm
+ \sum_n \frac{1}{\sqrt{2}} W_\mu^{(n)} \Big\{
g_{tb,L}^{W^{(n)}} \bar{b}_L \gamma^\mu t_L
+ g_{tb,R}^{W^{(n)}} \bar{b}_R \gamma^\mu t_R
+ g_{\tau,L}^{W^{(n)}} \bar{\tau}_L \gamma^\mu \nu_{\tau L}
+ g_{\tau,R}^{W^{(n)}} \bar{\tau}_R \gamma^\mu \nu_{\tau R} \Big\} 
+ \hbox{h.c.} \cr
\noalign{\kern 10pt}
&&\hskip -1.4cm
+\sum_n  \frac{1}{\cos \theta_W}   Z_\mu^{(n)}
\Big\{  g_{t L}^{Z^{(n)}}  \bar{t}_L \gamma^\mu t_L
+ g_{t R}^{Z^{(n)}} \bar{t}_R \gamma^\mu t_R
+ g_{b L}^{Z^{(n)}} \bar{b}_L \gamma^\mu b_L
 + g_{b R}^{Z^{(n)}} \bar{b}_R \gamma^\mu b_R  \cr
\noalign{\kern 10pt}
&&\hskip 1.cm
+  g_{\nu_\tau  L}^{Z^{(n)}}  \bar{\nu}_{\tau L} \gamma^\mu \nu_{\tau L}
+  g_{\nu_\tau  R}^{Z^{(n)}}  \bar{\nu}_{\tau R} \gamma^\mu \nu_{\tau R}
+ g_{\tau L}^{Z^{(n)}} \bar{\tau}_L \gamma^\mu \tau_L
 + g_{\tau R}^{Z^{(n)}} \bar{\tau}_R \gamma^\mu \tau_R \Big\}\cr
\noalign{\kern 10pt}
&&\hskip -1.4cm
+ \sum_n  G_\mu^{(n)a} 
\Big\{ (g_{t L}^{G^{(n)}} \bar{t}_L \gamma^\mu \onehalf \lambda^a t_L
+g_{t R}^{G^{(n)}} \bar{t}_R \gamma^\mu \onehalf \lambda^a t_R) \cr
\noalign{\kern 5pt}
&&\hskip 4.cm
+ (g_{b L}^{G^{(n)}} \bar{b}_L \gamma^\mu  \onehalf \lambda^a b_L
+ g_{b_R}^{G^{(n)}} \bar{b}_R \gamma^\mu  \onehalf \lambda^a b_R) \Big\}.
\label{Gcoupling1}
\eeqn
From the $H$ parity invariance the $W'$, $Z'$ and $A^{\hat 4}$ gauge boson towers 
do not couple to the quarks and leptons. 

The couplings of the photon tower with the $t$ and $b$ quarks and $\tau$ lepton 
are given, 
with $h_{\gamma^{(n)}}^L=h_{\gamma^{(n)}}^R 
= (g_B/g_A) h_{\gamma^{(n)}}^B \equiv h_{\gamma^{(n)}} (z)$,   
by
\beqn
&&\hskip -1.cm
g_{t L}^{\gamma^{(n)}} =
g_A \int_1^{z_L} dz  \,  h_{\gamma^{(n)}}
\Big\{  (a_U^2 +a_{B+t}^2 ) C_L(\lambda_t)^2
    + a_{t'}^2 S_L(\lambda_t)^2 \Big\} , \cr
\noalign{\kern 7pt}
&&\hskip -1.cm
g_{b L}^{\gamma^{(n)}} =
g_A \int_1^{z_L} dz \,  h_{\gamma^{(n)}} 
\Big\{  (a_b^2 +a_{D+X}^2 ) C_L(\lambda_b)^2
    + a_{b'}^2 S_L(\lambda_b)^2  \Big\}, \cr
\noalign{\kern 7pt}
&&\hskip -1.cm
g_{\tau L}^{\gamma^{(n)}} =
g_A \int_1^{z_L} dz \,  h_{\gamma^{(n)}}
\Big\{ (a_{L_{3Y}}^2 +a_{\tau + L_{1X}}^2 ) C_L(\lambda_\tau)^2  
+ a_{\tau'}^2 S_L( \lambda_\tau)^2  \Big\}.
\label{kkphoton}
\eeqn
Here $C_L(\lambda_t) = C_L(z; \lambda_t, c_t)$ etc..
The formulas for  right-handed fermions are obtained
from those for the corresponding left-handed fermions 
by replacing $C_L $ and $S_L $ by $S_R $ and $C_R $, respectively.
The couplings of the $W$ boson towers are given by
\beqn
&&\hskip -1.cm
g_{tb,L}^{W^{(n)}} =
%\frac{g_A}{\sqrt{2 r_{W^{(n)}}}}   
g_A \int_1^{z_L} dz  \,  \Big\{ 
2 h_{W^{(n)}} (a_b a_{B+t} +a_U a_{D+X} ) C_L(\lambda_t)C_L(\lambda_b) \cr
\noalign{\kern 7pt}
&&\hskip 1.cm
+ \sqrt{2} h_{W^{(n)}}^\wedge \big(a_B a_{t'} C_L(\lambda_b) S_L(\lambda_t)
 - a_{b'} a_U S_L(\lambda_b) C_L(\lambda_t) \big)  \Big\} ~, \cr
\noalign{\kern 7pt}
&&\hskip -1.cm
g_{\tau,L}^{W^{(n)}} =
g_A \int_1^{z_L} dz  \,  \Big\{ 
2 h_{W^{(n)}} (a_{\nu_\tau} a_{\tau + L_{1X}} +a_{L_{3Y}} a_{L_{2Y}+L_{3X}} ) C_L(\lambda_\tau)C_L(\lambda_{\nu_\tau}) \cr
\noalign{\kern 7pt}
&&\hskip 1.cm
+ \sqrt{2} h_{W^{(n)}}^\wedge \big(a_{L_{3Y}} a_{\nu_\tau '} C_L(\lambda_\tau) S_L(\lambda_{\nu_\tau})
 - a_{\tau'} a_{\nu_\tau} S_L(\lambda_{\nu_\tau}) C_L(\lambda_\tau) \big)  \Big\} ~
\label{kkW}
\eeqn
where $h_{W^{(n)}} \equiv h_{W^{(n)}}^L = h_{W^{(n)}}^R$.  
The couplings of the $Z$ boson towers are parametrized as
\beqn
&&\hskip -1.cm
g^{Z^{(n)}}_{tL,R}= + \frac{1}{2}g^{Z^{(n)}, T}_{tL,R}
- \frac{2}{3} g^{Z^{(n)}, Q}_{tL, R} \sin^2 \theta_W ~, \cr
\noalign{\kern 10pt}
&&\hskip -1.cm
g^{Z^{(n)}}_{bL,R}= - \frac{1}{2}g^{Z^{(n)}, T}_{bL,R}
+ \frac{1}{3} g^{Z^{(n)}, Q}_{bL, R} \sin^2 \theta_W ~, \cr
\noalign{\kern 10pt}
&&\hskip -1.cm
g^{Z^{(n)}}_{\nu_\tau L,R}= + \frac{1}{2}g^{Z^{(n)}, T}_{\nu_\tau L,R} ~, \cr
\noalign{\kern 10pt}
&&\hskip -1.cm
g^{Z^{(n)}}_{\tau L,R}= - \frac{1}{2}g^{Z^{(n)}, T}_{\tau L,R}
+  g^{Z^{(n)}, Q}_{\tau L, R} \sin^2 \theta_W ~.
\label{KKZ1}
\eeqn
In the SM $g^{Z, T}_{f L} = g^{Z, Q}_{f L} = g^{Z, Q}_{f R} = g_w$ 
and $g^{Z, T}_{f R} =0$.
In the current model, with the aid of (\ref{Zboson1}), one finds that
\beqn
&&\hskip -1.cm
g^{Z^{(n)}, T}_{tL} = \frac{\sqrt{2} g_A}{\sqrt{r_{Z^{(n)}}}} \int_1^{z_L} dz
\Big\{ a_U^2 C_{Z^{(n)}}  C_L(\lambda_t)^2 
- 2 a_{B+t} a_{t'} \hat S_{Z^{(n)}}  C_L(\lambda_t) S_L (\lambda_t) \Big\} ~, \cr
\noalign{\kern 10pt}
&&\hskip -1.cm
g^{Z^{(n)}, T}_{bL} = \frac{\sqrt{2} g_A}{\sqrt{r_{Z^{(n)}}}} \int_1^{z_L} dz
\Big\{ a_b^2 C_{Z^{(n)}}  C_L(\lambda_b)^2 
+ 2 a_{D+X} a_{b'} \hat S_{Z^{(n)}}  C_L(\lambda_b) S_L (\lambda_b) \Big\} ~, \cr
\noalign{\kern 10pt}
&&\hskip -1.cm
g^{Z^{(n)}, T}_{\nu_\tau L} = \frac{\sqrt{2} g_A}{\sqrt{r_{Z^{(n)}}}} \int_1^{z_L} dz
\Big\{ a_{\nu_\tau}^2 C_{Z^{(n)}}  C_L(\lambda_{\nu_\tau})^2 
- 2 a_{L_{2Y}+L_{3X}} a_{\nu_\tau '} \hat S_{Z^{(n)}}  C_L(\lambda_{\nu_\tau})
     S_L (\lambda_{\nu_\tau}) \Big\} ~, \cr
\noalign{\kern 10pt}
&&\hskip -1.cm
g^{Z^{(n)}, T}_{\tau L} = \frac{\sqrt{2} g_A}{\sqrt{r_{Z^{(n)}}}} \int_1^{z_L} dz
\Big\{ a_{L_{3Y}}^2 C_{Z^{(n)}}  C_L(\lambda_\tau)^2 
+ 2 a_{\tau + L_{1X}} a_{\tau '} \hat S_{Z^{(n)}}  C_L(\lambda_\tau) 
S_L (\lambda_\tau) \Big\} ~, \cr
\noalign{\kern 10pt}
&&\hskip -1.cm
g^{Z^{(n)}, Q}_{tL} = \frac{\sqrt{2} g_A}{\sqrt{r_{Z^{(n)}}}} \int_1^{z_L} dz \, 
C_{Z^{(n)}} \Big\{ (a_U^2 + a_{B+t}^2 ) C_L(\lambda_t)^2 
+ a_{t'}^2 S_L (\lambda_t)^2 \Big\} ~, \cr
\noalign{\kern 10pt}
&&\hskip -1.cm
g^{Z^{(n)}, Q}_{bL} = \frac{\sqrt{2} g_A}{\sqrt{r_{Z^{(n)}}}} \int_1^{z_L} dz \, 
C_{Z^{(n)}} \Big\{ (a_b^2 + a_{D+X}^2 ) C_L(\lambda_b)^2 
+ a_{b'}^2 S_L (\lambda_b)^2 \Big\} ~, \cr
\noalign{\kern 10pt}
&&\hskip -1.cm
g^{Z^{(n)}, Q}_{\tau L} = \frac{\sqrt{2} g_A}{\sqrt{r_{Z^{(n)}}}} \int_1^{z_L} dz \, 
C_{Z^{(n)}} \Big\{ (a_{L_{3Y}}^2 + a_{\tau+L_{1X}}^2 ) C_L(\lambda_\tau)^2 
+ a_{\tau '}^2 S_L (\lambda_\tau)^2 \Big\} ~,
\label{KKZ2}
\eeqn
where $C_{Z^{(n)}} = C(z; \lambda_{Z^{(n)}})$ etc..

The  couplings of the gluon tower $g_{t I}^{G^{(n)}}$ and $g_{b I}^{G^{(n)}}$
are obtained from the photon tower couplings $g_{t I}^{\gamma^{(n)}}$ and 
$g_{b I}^{\gamma^{(n)}}$ with the replacement of
the five-dimensional coupling,
$g_{t I}^{G^{(n)}} = (g_C/g_A) g_{t I}^{\gamma^{(n)}}$
and $g_{b I}^{G^{(n)}} = (g_C/g_A) g_{b I}^{\gamma^{(n)}}$.
The photon and gluon couplings are universal, that is, 
$e= g_{tI}^{\gamma^{(0)}}=g_{bI}^{\gamma^{(0)}} =(g_A/\sqrt{L})\sin\theta_W$.
The other couplings exhibit violation of the universality as evaluated below.

\subsection{Zero mode couplings}

The numerical values for the various gauge couplings are obtained with the input
parameters given in Sec.~2.
The couplings of $W$ boson with quarks and leptons 
are tabulated in Tables~\ref{tab:wq}.
\begin{table}[htb]
\begin{center}
\caption{The couplings of $W$ boson with quarks and leptons, 
$g_{f}^{(W)}\sqrt{L}/g_A$.}
\label{tab:wq}
\vskip 5pt
\begin{tabular}{|c|ccc|ccc|}
\hline
$z_L$ 
& $u_L d_L$
& $c_L s_L$ 
& $t_L b_L$
& $\nu_{e L}  e_L $
& $\nu_{\mu L} \mu_L$
& $\nu_{\tau L}  \tau_L$
\\
\hline
$10^{15}$ 
&
1.0053 
&
1.0053
& 
0.9816
&
1.0053
&
1.0053 
& 
1.0053
\\
$10^{10}$ 
& 
1.0079
& 
1.0079 
& 
0.9730
& 
1.0079
&
1.0079 
&
1.0079
\\
$10^5$ 
& 
1.0154
& 
1.0154
& 
0.9470
& 
1.0154 
& 
1.0154 
& 
1.0153
\\
\hline
\noalign{\kern 7pt}
\hline
$z_L$ 
& $u_R d_R$
& $c_R s_R$
& $t_R b_R$ 
& $\nu_{eR} e_R$
& $\nu_{\mu R} \mu_R$
& $\nu_{\tau R} \tau_R$
\\
\hline
$10^{15}$ 
&$-5 \times 10^{-12}$ 
& $-5 \times 10^{-8}$ 
& 
$-0.0009$
& $-3\times 10^{-22}$ 
& $-4\times 10^{-16}$ 
& $-6\times 10^{-17}$ 
\\
$10^{10}$ 
& $-6 \times 10^{-12}$  
& $-7 \times 10^{-8}$ 
& 
$-0.0014$
& $-4\times 10^{-22}$ 
& $-7\times 10^{-16}$ 
& $-2\times 10^{-13}$ 
\\
$10^5$ 
& $-9 \times 10^{-12}$ 
& $-1 \times 10^{-7}$ 
& 
$-0.0031$
& $-5\times 10^{-22}$ 
& $-1\times 10^{-18}$ 
& $-2\times 10^{-16}$ 
\\
\hline
\end{tabular}
\end{center}
\end{table}
The ratios of the couplings to the 4D $SU(2)$ coupling,  $g_{f}^{(W)}\sqrt{L}/g_A$,
have been tabulated.  Except for $tb$, the couplings are almost universal.
For the $t_L b_L$ coupling the deviation amounts to $2 \sim 6\,$\% for 
$z_L= 10^{15} \sim 10^5$. The $t_R b_R$ coupling is about 
$0.09 \sim 0.3\,$\% of the left-handed coupling for $z_L= 10^{15} \sim 10^5$. 

The couplings of $Z$ boson with  quarks are tabulated in Tables~\ref{tab:zql}.
For reference,  the tree-level values in  the standard model,
$1/2-(2/3)\sin^2\theta_W$ and $-1/2+(1/3)\sin^2\theta_W$  for left-handed 
quarks and $-(2/3)\sin^2\theta_W$ and $(1/3)\sin^2\theta_W$  for   
right-handed quarks are also listed. 
As we shall see below, the small violation of the universality gives
a better fit to the forward-backward asymmetry data.
As a general character for left-handed and right-handed quarks,
it is found that the coupling of
right-handed quarks for a small warp factor
tends to deviate from the standard model values.

\begin{table}[htb]
\begin{center}
\caption{The couplings of $Z$ boson with  quarks,
$g_{f}^{(Z)}\sqrt{L}/g_A$.}
\label{tab:zql}
\vskip 5pt
\begin{tabular}{|c|ccc|ccc|}
\hline 
$z_L$  
& $u_L$ 
& $c_L$
& $t_L$
& $d_L$
& $s_L$
& $b_L$ \\
\hline
$10^{15}$  
& 
0.3485
& 
0.3485
& 
0.3219
&
$-0.4260$
& 
$-0.4260$
&
$-0.4265$
\\
$10^{10}$ 
& 
0.3501
& 
0.3501
& 
0.3086
& 
$-0.4276$
& 
$-0.4276$
& 
$-0.4288$
\\
$10^5$ 
& 
0.3548
& 
0.3548
& 
0.2558
& 
$-0.4325$
& 
$-0.4325$
& 
$-0.4369$
\\
\hline
SM
&\multicolumn{3}{c|}{
0.3459
} &\multicolumn{3}{c|}{
$-0.4229$
}
\\
\hline
\noalign{\kern 5pt}
\hline 
$z_L$ 
& $u_R$
& $c_R$
& $t_R$
& $d_R$
& $s_R$
& $b_R$ 
\\
\hline
$10^{15}$ 
& 
$-0.1562$
& 
$-0.1562$
& 
$-0.1835$
& 
0.07811
& 
0.07809
& 
0.07806
\\
$10^{10}$ 
& 
$-0.1570$
& 
$-0.1570$
& 
$-0.2002$
&
0.07852
& 
0.07847
& 
0.07839
\\
$10^5$ 
& 
$-0.1595$
& 
$-0.1593$
& 
$-0.2656$
& 
0.07976
& 
0.07965
& 
0.07928
\\
\hline
SM
&\multicolumn{3}{c|}{
$-0.1541$
} &\multicolumn{3}{c|}{
0.07707
}
\\
\hline
\end{tabular}
\end{center}
\end{table}

The couplings of $Z$ bosons with leptons are tabulated in Table~\ref{tab:zl}.
They are not very sensitive to the generation.
As general tendency, the couplings deviate more from those in the standard model 
as  the warp factor becomes smaller.  
\begin{table}[htb]
\begin{center}
\caption{The couplings of $Z$ bosons with leptons, $g_{f}^{(Z)}\sqrt{L}/g_A$. }
\label{tab:zl}
\vskip 5pt
\begin{tabular}{|c|ccc|ccc|}
\hline 
$z_L$ 
& $e_L$
& $\mu_L$
& $\tau_L$
& $e_R$
& $\mu_R$
& $\tau_R$
\\ \hline
$10^{15}$ 
& 
$-0.2710$
& 
$-0.2710$
& 
$-0.2710$
& 
0.2344
& 
0.2343
& 
0.2343
\\
$10^{10}$ 
& 
$-0.2725$
& 
$-0.2725$
& 
$-0.2725$
& 
0.2356
& 
0.2355
& 
0.2354
\\
$10^5$  
& 
$-0.2771$
& 
$-0.2771$
& 
$-0.2771$
& 
0.2394
& 
0.2391
& 
0.2389
\\
\hline
SM
&\multicolumn{3}{c|}{
$-0.2688$
} &\multicolumn{3}{c|}{0.2312}
 \\
\hline
\noalign{\kern 5pt}
\hline 
$z_L$  
& $\nu_{eL}$
& $\nu_{\mu L}$
& $\nu_{\tau L}$
& $\nu_{eR}$
& $\nu_{\mu R}$
& $\nu_{\tau R}$
\\ \hline
$10^{15}$ 
& 
0.5035
& 
0.5035
& 
0.5035
& $-1.4\times 10^{-13}$
& $-7.2\times 10^{-9}$ 
& $-2.3\times 10^{-6}$ 
\\ 
$10^{10}$ 
& 
0.5052
& 
0.5052
& 
0.5052
& $-1.8\times 10^{-13}$
& $-9.7\times 10^{-9}$ 
& $-3.2\times 10^{-6}$ 
\\
$10^5$ 
& 
0.5102
& 
0.5102
& 
0.5101
& $-2.5\times 10^{-13}$ 
& $-1.5\times 10^{-8}$ 
& $-5.4\times 10^{-6}$ 
\\
\hline
SM
&\multicolumn{3}{c|}{0.5} 
&\multicolumn{3}{c|}{0}
\\
\hline
\end{tabular}
\end{center}
\end{table}

In the standard model the couplings of $Z$ boson with fermions are 
described by the weak coupling and their quantum number, namely by
$(g_w/\cos\theta_W)(T^3 -Q \sin^2\theta_W)$, at the tree level.
In the present model they have an analogous form given 
by $(1/\cos\theta_W)(g_{T,L} T^3 - g_{Q,L} Q \sin^2\theta_W)$ for left-handed fermions and
by $(1/\cos\theta_W)(g_{T,R}  - g_{Q,R} Q \sin^2\theta_W)$ for right-handed fermions.
Here $g_T$ and $g_Q$ depends on the  flavor of fermions.
It is found that $g_{T,L} \approx g_{Q,L}$.
For right-handed fermions  the absolute value of  $g_{T,R}\sqrt{L}/g_A$ are 
small for  $t$-quark ($\siml 10^{-2}$)
and very small for the others ($\siml  10^{-6}$),
but $g_{Q,R}\sqrt{L}/g_A$ can be of order  ${\cal O}(1)$ 
which leads to  deviation from the standard model.
The couplings of $Z$ boson with right-handed neutrinos are very
small as neutral fields have only the $g_T$ component.
For a similar reason the couplings of KK $Z$ boson with right-handed neutrinos 
turn out very small.

\subsection{Forward-backward asymmetry}

The forward-backward asymmetry on the $Z$ resonance
is given by
\beeq
A_{FB}^f = \frac{3}{4} \, \left[
\frac {(g_{e L}^{Z})^2  -(g_{e R}^{Z})^2}{(g_{e L}^{Z})^2  + (g_{e R}^{Z})^2 }  \right]
\, \left[
\frac {(g_{f L}^{Z})^2 -(g_{f R}^{Z})^2}{ (g_{f L}^{Z})^2 + (g_{f R}^{Z})^2 } \right] ,
\label{AFB1}
\eneq
which is evaluated from the gauge couplings given in the preceding subsection.
$A_{FB}^f$ does not depend on the absolute common magnitude of $g_A$,
but sensitively depends on $\sin^2 \theta_W$.
The branching fractions of various decay modes of the $Z$ boson also
sensitively depend on $\sin^2 \theta_W$.
We have determined the value of $\sin^2 \theta_W$ to minimize $\chi^2$ of
those experimental data as tabulated in Table \ref{chi2fit}.
The value of $\sin^2 \theta_W$ turns out a bit smaller than that in
the standard model.

With given $\sin^2 \theta_W$ the numerical values of $A_{FB}^f$  are shown in 
Table~\ref{tab:afb}.\footnote{The result for $z_L=10^{15}$ has been given
in Ref.~\cite{Uekusa2009}.  A slight difference in the
numerical values is due to the different choice 
of the values of the input parameters.}
The experimental values are quoted from 
Ref.~\cite{Amsler:2008zzb}.
The current model gives rather good fit for the forward-backward asymmetries
$A_{FB}^f$, though the fit to the $Z$ decay
fractions becomes poor for smaller values
of $z_L$.
\begin{table}[htb]
\begin{center}
\caption{$\chi^2$ fit for $A_{FB}$ and  $Z$ decay 
fractions.
The values of $m_\KK$, $m_H$ and $m_W^{\rm tree}$
 ($W$ mass at the tree level) are also listed.
}
\label{chi2fit}
\vskip 10pt
\begin{tabular}{|c||c|c|c|c||c|}
\hline
&\# of data & $z_L=10^{15}$ & $10^{10}$ & $10^5$ &SM \\
\hline
$\sin^2 \theta_W$ &&0.2309 & 0.2303 & 0.2284 & 0.2312 \\
\hline
$\chi^2$ [$A_{FB}$] & 6 
& 
6.3
& 
6.4
&
7.1
&
10.8
\\
$\chi^2$ [$Z$ decay fractions] & 8
& 
16.5
& 
37.7
& 
184.5
& 
13.6
\\
\hline 
Sum of two $\chi^2$ & 14
& 
22.8
&
44.1
& 
191.6
& 
24.5
\\
\hline 
$m_\KK$ (GeV) &  & 1466 & 1193 & 836 &\\
$m_H$ (GeV) &  &135 &108 &72 &\\
$m_W^{\rm tree}$ (GeV) &  &79.84 &79.80 &79.71 &79.95 \\
\hline
\end{tabular}
\end{center}
\end{table}
\begin{table}[htb]
\begin{center}
\caption{The forward-backward asymmetry on the $Z$ resonance,
$A_{FB}^f$. \label{tab:afb}}
\vskip 5pt
\begin{tabular}{|c|c|c|c|c|c|}
\hline
& Exp. & $z_L =10^{15}$
& $z_L =10^{10}$ & $z_L = 10^5$ &SM \\
\hline
$e$ &
$0.0145 \pm 0.0025$ 
& 
0.0156
& 
0.0157
& 
0.0159 
& $0.01633 \pm 0.00021$
\\
$\mu$ &
$0.0169 \pm 0.0013$ 
& 
0.0156
& 
0.0157
& 
0.0160 
&
\\
$\tau$ &
$0.0188 \pm 0.0017$ 
& 
0.0156
& 
0.0158
& 
0.0161
&
\\ \hline
$s$ &
$0.0976 \pm 0.0114$ 
& 0.1011 
& 
0.1014
& 
0.1019
& $0.1035 \pm 0.0007$
\\
$c$ &
$0.0707 \pm 0.0035$ 
& 
0.0720
& 
0.0721 
& 
0.0725 
& $0.0739 \pm 0.0005$
\\
$b$ &
$0.0992 \pm 0.0016$ 
& 
0.1011 
& 
0.1014 
& 
0.1021  
& $0.1034 \pm 0.0007$
\\
\hline
\end{tabular}
\end{center}
\end{table}

\subsection{Decay width}

The partial decay width of $Z$ boson is given by
\beqn
&&\hskip -1.cm
\Gamma( Z \to f\bar{f} )   =
\frac{m_Z}{12 \pi  \cos^2 \theta_W } 
F(  g_{f L}^{(Z)},   g_{f R}^{(Z)}, m_f, m_Z)   ~, \cr
\noalign{\kern 10pt}
&&\hskip -1.cm
F(g_{f L}, g_{f R}, m_f, m_V )
=  \bigg\{ \frac{(g_{f L})^2 + (g_{f R})^2}{2}
+ 2 g_{f L} g_{f R}   \frac{m_f^2}{m_{V}^2} \bigg\}
\sqrt{1- \frac{4 m_f^2}{m_V^2}}  ~.
\label{Zdecay1}
\eeqn
Here the couplings $g_{fL}^{(Z)}$ and $g_{fR}^{(Z)}$  are given
in Tables~\ref{tab:zql} and \ref{tab:zl}.
For quarks the formula should be multiplied by a factor $3 (1+ \alpha_s/\pi)$.

For $z_L=10^{15}, 10^{10}, 10^5$,  the branching fractions in $Z$ decay
are shown in Table~\ref{tab:decay}. The experimental values are quoted from 
Ref.~\cite{Amsler:2008zzb}.  The tree level prediction for branching fractions reproduces
the pattern of the experimental values well for $z_L=10^{15}$.

The total decay width $\Gamma_{\rm tot}$ depends on $\alpha(m_Z)$.
The value of $\alpha(m_Z)$ determined to fit the experimental value 
$\Gamma_{\rm tot}$  does not agree well with the value determined by
renormalization group from the low energy data.  
For $z_L = 10^{15}$, for instance, one finds $\alpha^{-1}(m_Z) = 130.5$.   
At the moment one cannot reliably evaluate one loop corrections to $\Gamma_{\rm tot}$
in the gauge-Higgs unification scenario and this mismatch is understood
within that error.

\begin{table}[htb]
\begin{center}
\caption{The branching fractions in the $Z$ boson decay.
The invisible decay in the model
means the decay into $\nu_e + \nu_\mu + \nu_\tau$.
}
\label{tab:decay}
\vskip 5pt
\begin{tabular}{|c|ccc|c|}
\hline
$z_L$ & $10^{15}$ & $10^{10}$ & $10^5$ & Exp.
\\ \hline 
$e~(\%)$ & 
3.374
& 
3.382
& 
3.403
& $3.363 \pm 0.004$ 
\\
$\mu~(\%)$ & 
3.373
& 
3.380 
& 
3.400
& $3.366 \pm 0.007$
\\
$\tau~(\%)$ & 
3.368
& 
3.374
& 
3.392
& $3.370 \pm 0.008$
\\ \hline
invisible $(\%)$ & 
19.99
& 
19.95
& 
19.82
& $20.00 \pm 0.06$
\\ \hline
$(u + c)/2~(\%)$ & 
11.93
& 
11.94
& 
11.95
& $11.6 \pm 0.6$
\\ 
$(d + s + b)/3 \, (\%)$ & 
15.34
& 
15.34
& 
15.36
& $15.6 \pm 0.4$
\\
$c ~(\%)$ & 
11.93
& 
11.94
& 
11.95
& $12.03 \pm 0.21$
\\
$b ~(\%)$ & 
15.34
& 
15.37
& 
15.53
& $15.21 \pm 0.05$ 
\\ 
\hline
\end{tabular}
\end{center}
\end{table}

\section{Production of Higgs bosons at colliders}

The mass of the Higgs boson is in the range 70 GeV - 140 GeV, depending
on the warp factor $z_L$.  Higgs bosons can be copiously produced at colliders 
at high energies.  
At $\theta_H = \onehalf \pi$, however, there emerges the $H$ parity 
conservation so that Higgs bosons can be produced only in pairs, 
provided no other KK modes of $P_H$ odd fields are produced.
Furthermore, the Higgs boson becomes stable at $\theta_H = \onehalf \pi$
so that conventional ways of detecting the Higgs boson, namely of finding
decay products of the Higgs boson, turns out fruitless.  
In the current scheme the produced Higgs boson appears as missing energy
and momentum.  At colliders there appear at least two particles of missing
energy and momentum, which makes detection hard.  
There is large background containing neutrinos. 

An interesting feature of the present model is that
the stable Higgs boson is much lighter than the KK particles,
that is, $m_H\ll m_{KK}$ as seen in Table \ref{chi2fit}.
Hence it is natural to investigate the Higgs production with 
the effective Lagrangian among low energy fields 
($W$, $Z$, quarks and leptons) at 
$\theta_H = \onehalf \pi$ \cite{HK,HKT},
\beeq 
{\cal L}_\eff  
 \sim - \Big\{ m_W^2 W_\mu^\dagger W^\mu 
     + \frac{1}{2} m_Z^2 Z_\mu Z^\mu \Big\}  \cos^2 \frac{H}{f_H} 
- \sum_a m_a \psibar_a \psi_a \cos \frac{H}{f_H} ~.
\label{effective1}
\eneq
Here $f_H \sim 246\,$GeV. The form of (\ref{effective1}) is valid only when
the relevant energy scale is sufficiently smaller than $m_\KK$.
For the pair production of Higgs bosons (\ref{effective1}) leads to
\beeq
{\cal L}_\eff   \sim
\sum_a  \frac{m_a}{2f_H^2} \, H^2 \psibar_a \psi_a
+ \frac{m_W^2}{f_H^2} H^2 W_\mu^\dagger W^\mu 
+\frac{m_Z^2}{2f_H^2} H^2 Z_\mu Z^\mu ~ .
\label{effective2}
\eneq
The sign of the $H^2W^\dagger W$ and $H^2 ZZ$  couplings is opposite
to that in SM.\cite{Sakamura1}
Collider signatures of Higgs bosons in the current model have been previously 
investigated with this effective Lagrangian 
by Cheung and Song\cite{Cheung2010a} and by Alves.\cite{Alves2010}

\subsection{Pair production of Higgs bosons at LHC}
Pair production processes of Higgs bosons at LHC have been 
studied in Refs.\ \cite{Cheung2010a} and \cite{Alves2010}. 
Cheung and Song evaluated the cross section of the Higgs pair production
associated with a $W$ or $Z$ boson and found that 
the $ZHH$($WHH$) cross section is $0.2(0.4)\ \mathrm{fb}$
for the case that $m_H=70\ \mathrm{GeV}$ and the missing transverse
momentum $\not\!p_T$ is larger than $100\ \mathrm{GeV}$.
On the other hand the cross section of the background process 
$ZZ\rightarrow Z\nu\bar\nu$ ($WZ\rightarrow W\nu\bar\nu$)
was estimated as $370(390)\ \mathrm{fb}$. Thus   positive 
identification of either of the signals is virtually impossible 
assuming an  integrated luminosity of $100\ \mathrm{fb}^{-1}$.

Alves studied the Higgs pair production in the weak boson fusion (WBF),
in which the signal is a pair of forward and backward jets and a missing
transverse momentum. This signal is quite similar to that of
the single production of the Higgs boson decaying invisibly \cite{EZ2000}.
In Ref.~\cite{Alves2010}, the signal cross section at 14 TeV LHC is estimated 
as $4.05(4.03)\ \mathrm{fb}$ for $m_H=70(90)\ \mathrm{GeV}$ using the
same set of cuts employed in Ref.~\cite{EZ2000}. The background cross section
is the same as that in Ref.~\cite{EZ2000} and amounts to $167\ \mathrm{fb}$.
Alves concluded that $255(257)\ \mathrm{fb}^{-1}$ is required for a $5\sigma$
discovery.

Here we present a brief estimate of the cross section of the Higgs pair 
production by the WBF in the present model by relating it to 
the single Higgs production by the WBF in the SM.
Inspecting the relevant Feynman rules in the present model
and the SM, we find that
$f_H^2|{\cal M}(HH)|^2=|{\cal M}(h)|^2_{m_h^2=m_{HH}^2}$,
where $h$ represents the Higgs boson in the SM,
${\cal M}(HH)$ (${\cal M}(h)$) denotes the amplitude of the double (single)
Higgs production by the WBF process in the present model (SM), 
$m_h$ is the Higgs boson mass in the SM,
$m_{HH}$ is the invariant mass of the pair of Higgs bosons
in the present model. Taking the two-body phase space of the Higgs pair
and the statistical factor due to the existence of identical particles into
account, we obtain the following relation,
\begin{equation}
 \frac{d\sigma(HH)}{dm_{HH}^2}=
  \frac{1}{32\pi^2 f_H^2}
  \sqrt{1- \frac{4m_H^2}{m_{HH}^2} } ~  \sigma(h) \Big|_{m_h^2=m_{HH}^2},
  \label{dSigWBF}
\end{equation}
where $\sigma(HH)$ ($\sigma(h)$) represents the cross section of
the double (single) Higgs production by the WBF in the present model (SM).

The total cross section is evaluated by integrating Eq.~(\ref{dSigWBF}) 
over $m_{HH}^2$ in the kinematically allowed interval.
The upper value of the integration region could be as large
as the center of mass energy squared of $pp$ collisions in principle. 
However, as mentioned above, the effective Lagrangian in 
Eq.~(\ref{effective2}) is applicable in a limited energy scale.
We choose $4 m_{KK}^2$ as the upper value for an illustration. We note that
$4 m_{KK}^2\simeq (1.7\ \mathrm{TeV})^2$ for the case that $z_L=10^5$
is approximately the same as the unitarity bound of $1.8\ \mathrm{TeV}$
obtained in Ref.~\cite{Alves2010}.
We evaluate the right-hand side of Eq.~(\ref{dSigWBF}) at the parton level
with CTEQ6L parton distribution function \cite{CTEQ6L} and the cuts used in 
Refs.~\cite{Alves2010,EZ2000}. Our numerical calculation is done by 
MadGraph/MadEvent \cite{MGMEV4} without hadronization or detector
simulation.

The signal cross section at $14\ \mathrm{TeV}$ LHC is
approximately $1.3\ \mathrm{fb}$ for $z_L=10^5-10^{15}$.
Our result is smaller by about a factor of 3 than that of 
Ref.~\cite{Alves2010}. Thus, an integrated luminosity of a few 
$\mathrm{ab}^{-1}$ seems to be required to observe the signal.

\subsection{Pair production of Higgs bosons at ILC}
\begin{figure}[htb]
\begin{center} 
\includegraphics[width=5cm]{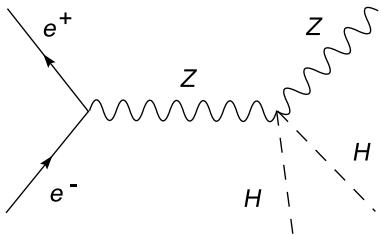}
\caption{Pair production of Higgs bosons at ILC.}
\label{fig:eezhh}
\end{center}
\end{figure}

Cheung and Song have studied the Higgs pair production process 
$e^- e^+ \to Z HH$ at ILC along with the background process 
$e^- e^+ \to Z\nu\bar\nu$. The Feynman diagram of the signal
process is depicted in Fig.~\ref{fig:eezhh}.
The integrated luminosity for a $5\sigma$
discovery seems to be larger than several $\mathrm{ab}^{-1}$ at 
$500\ \mathrm{GeV}$ ILC according to their result.

The differential cross section of $e_R^- e_L^+ \to Z_T HH$,
where $Z_T$ denotes the transversely-polarized $Z$ boson, 
is given by
\bea
\frac{d\sigma_{RL}^T}{dx \, d\cos \theta }
=\frac{g_R^2 \, m_Z^4 \, s }{ 2(4\pi)^3  f_H^4  (s-m_Z^2)^2}
\sqrt{ \frac{(x_{\rm max}-x) (x^2 -x_{\rm min}^2)}{1+ x_{\rm min}^2 /4 -x}}
~ {1+\cos^2\theta\over 2} ~.
\label{Xsection1}
\eea
Here  $g_R$ denotes the coupling constant of the right-handed electron
to the $Z$ boson, which is given by 
$g_R \big|_{SM} = -\sqrt{g^2 +g^{'2}} \, \sin^2 \theta_W$
in the standard model,
and $\theta$ is the angle between the momentum of the electron and that 
of the $Z$ boson in the center-of-mass system.
The energy of the $Z$ boson normalized to the beam energy, $x$, and
its minimal and maximal values are given by 
\beeq
x = \frac{E_Z}{\sqrt{s}/2} ~,
\quad
x_{\rm min} = \frac{m_Z}{\sqrt{s}/2} ~,
\quad
x_{\rm max} =1- \frac{4 m_H^2}{s} +\frac{x_{\rm min}^2}{4} ~ .
\eneq

For $e_R^- e_L^+ \to  Z_L HH$, where $Z_L$ denotes the longitudinal $Z$ boson,
the differential cross section is given by
\beeq
\frac{d\sigma_{RL}^L}{dx \, d\cos \theta }
=\frac{g_R^2 \, m_Z^4 \, s }{ 2(4\pi)^3  f_H^4  (s-m_Z^2)^2}
\sqrt{ \frac{(x_{\rm max}-x) (x^2 -x_{\rm min}^2)}{1+ x_{\rm min}^2 /4 -x}}
~\frac{x^2}{x_{\rm min}^2}\,  \frac{1-\cos^2 \theta}{2} ~.
\label{Xsection2}
\eneq
For the case of $e_L^- e_R^+$,  $g_R$ should be replaced by
$g_L$ in the above formulas.
In the standard model $g_L \big|_{SM}=\sqrt{g^2 +{g'}^2} \,  (1/2-\sin^2\theta_W)$. 

\begin{figure}[htb]
\begin{center} 
\vskip 10pt
\includegraphics[height=6.cm]{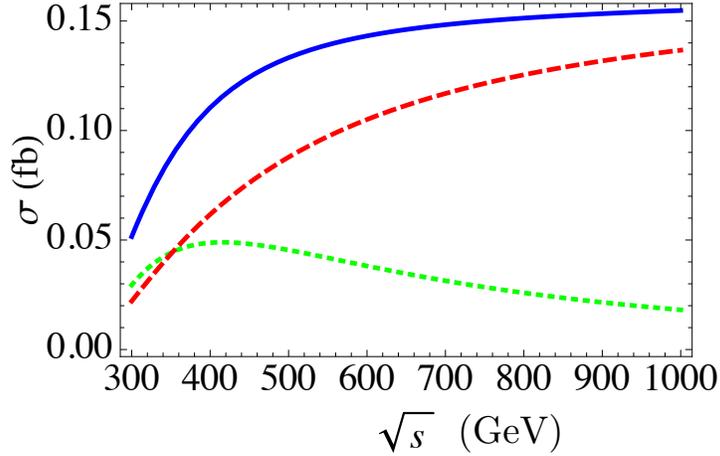}
\caption{Cross sections of Higgs pair production at ILC for
         $z_L=10^5$. 
         The dotted (green) line shows the $Z_T$ mode, 
         the dashed (red) line is the $Z_L$ mode and
         the solid (blue) represents their sum.}
\label{fig:ILCcs}
\end{center}
\end{figure}

Figure \ref{fig:ILCcs} shows the total cross sections of
$e^- e^+\to Z_T H H$ and $e^- e^+\to Z_L H H$ and their sum
as functions of $\sqrt{s}$ for 
$z_L =10^5$. As $\sqrt{s}$ increases, the cross section of 
the $Z_L$ mode asymptotically becomes constant, violating the unitarity
bound. This is expected because the low energy effective Lagrangian in 
Eq.~(\ref{effective1}) does not contain  vertices with
an odd number of Higgs fields after integrating out all heavy KK modes. 
With adding diagrams with such vertices to the one in Fig.~\ref{fig:eezhh},
the leading terms in the amplitudes cancel among each other in the standard 
model so that the unitarity is maintained.
In the present model the KK  modes  appearing  as internal lines of 
the relevant diagrams are supposed to rescue the unitarity. 
Put it differently, the effective Lagrangian is applicable only for 
the case that contributions of the KK modes are negligible, that is, 
when $\sqrt{s}\ll m_{KK}$. Accordingly, unless otherwise stated, 
we take $\sqrt{s}= 500 \,$GeV in the following numerical calculation 
in this subsection.

The major background is $e^- e^+ \to Z \nu_\alpha \bar{\nu}_\alpha$ 
($\alpha=e,\mu,\tau$).
Their total cross section is about
300\,fb for $M_\mathrm{mis} \geq 120\,$GeV, where $M_\mathrm{mis}$
is the invariant mass of the neutrino pair. The background is dominated by 
the electron neutrino mode due to the t-channel $W$ boson exchange. 
In order to reduce this large background, one may use beam polarizations.
We consider the limiting case of the purely right-handed electron and the purely
left-handed positron as an ideal case. 
As well as the beam polarizations,
the missing mass cut reduces the background. 
Corresponding to $m_H=72,\ 108,\ 135$ GeV for $z_L=10^5,\ 10^{10},\ 10^{15}$
respectively,  we take $M_\mathrm{mis}>120,\ 200,\ 250$ GeV.
We employ the GRACE system version 2 \cite{GRACE} in our numerical 
estimation of the background.

The statistical signification of the signal is defined by
\beeq
S = \frac{N_{\rm signal}}{\sqrt{N_{\rm signal} + N_{\rm BG}}} ~ ,
\label{significance1}
\eneq
where $N_{\textrm{\scriptsize signal}}$ and $N_{\textrm{\scriptsize BG}}$ are
the expected numbers of signal and background events,  respectively.
They are given as 
$N_\mathrm{signal(BG)}=L\,\sigma_\mathrm{signal(BG)}\,B_\mathrm{ev}$, where
$L$ is the integrated luminosity and
$B_{\rm ev} =1-B_{\rm  invisible}-B_{\tau\bar{\tau}}=1-0.200-0.037 =0.763$
is the effective visible branching ratio of the $Z$ boson.
The significance of the $Z_L$ mode turns out to be much larger than 
that of the $Z_T$ mode and we concentrate on the former in order to
evaluate the lower bound of integrated luminosity to establish the signal.
Employing $|\cos\theta|< 0.6$, which approximately maximizes the significance,
we obtain the significance of $e_R^-e_L^+\to Z_L HH$ as 
$S/\sqrt{L}=0.14,\ 0.073,\ 0.034$ for $z_L=10^5,\ 10^{10},\ 10^{15}$ 
respectively, where $L$ is in the fb$^{-1}$ unit. 
Thus, in order for $5\sigma$,
we need at least $1.3,\ 4.7,\ 21\,\textrm{ab}^{-1}$ for 
$z_L=10^5,\ 10^{10},\ 10^{15}$ respectively.
Since the KK mass scale rather high as
$m_{KK}=1466\,\mathrm{GeV}$ in the case of $z_L=10^{15}$, 
one can apply the effective Lagrangian to a higher energy.
For instance, we obtain $S/\sqrt{L}=0.11$ for $\sqrt{s}=750\,\mathrm{GeV}$,
and the required luminosity is $L>2.0\,\textrm{ab}^{-1}$.

\section{Spectrum of KK states}

One of the direct ways to see the extra dimension is to produce
KK excited modes of various particles and observe their decays.
In the current model the $H$ parity is conserved so that 
$P_H$-odd KK modes can be produced in a pair.  Production of 
a single KK mode occurs only for $P_H$-even modes.
In this section we determine spectra of various KK modes.

\subsection{KK gauge bosons}

The spectrum of KK gluons $G^{(n)}$ is identical to the spectrum of KK 
photons 
$A^{\gamma (n)}$. 
They are determined by the fourth equation in Eq.~(\ref{spectrum1}).
The masses of KK $W$ and $Z$ bosons, $W^{(n)}$ and $Z^{(n)}$, are 
determined  by the first and second equations in Eq.~(\ref{spectrum1}),
whereas those of $W^{\prime (n)}$, $Z^{\prime (n)}$ and $A^{\hat{4} (n)}$
are determined by the third and fifth equations.
The numerical values of the masses of the first five KK modes  are 
given in Table~\ref{KKgaugeboson1}.

We observe that among $P_H$-even modes
\beqn
&&\hskip -.5cm
m_{Z^{(1)}} < m_{W^{(1)}} < m_{G^{(1)}} < m_{W^{(2)}}  < m_{Z^{(2)}} \cr
\noalign{\kern 5pt}
&&\hskip -1.cm
< m_{Z^{(3)}}  < m_{W^{(3)}} < m_{G^{(2)}}  < m_{W^{(4)}}  < m_{Z^{(4)}} \cr
\noalign{\kern 5pt}
&&\hskip -1.cm
< m_{Z^{(5)}}  < m_{W^{(5)}} < m_{G^{(3)}}   < m_{W^{(6)}}  < m_{Z^{(6)}} < \cdots
\label{masspatern1}
\eeqn
irrespective of $z_L$.  Lighter  the  $n=0$ mode is, the  $n=1$ mode becomes heavier.
Masses of $P_H$-odd gauge bosons obey the pattern 
$m_{{W'}^{(n)}} = m_{{Z'}^{(n)}} \sim m_{{A}^{\gamma (n)}}$ and 
$m_{{A}^{\hat 4 (n)}} \sim m_{{W}^{(2n)}}$.

\begin{table}[htb]
\begin{center}
\caption{Mass spectra of KK gauge bosons  in unit of GeV.}
\label{KKgaugeboson1}
\vskip 5pt
\begin{tabular}{|c|ccccc|}
\hline 
&\multicolumn{5}{c|}{$%\myred{
A^{\gamma (n)}%}
 ~,~  G^{(n)}$}
\\
\hline
$z_L$ $\backslash$ $n$ &  1  & 2 & 3 & 4 & 5
\\
\hline
$10^{15}$ 
& 
1144
& 
2598
& 
4061
& 
5522
& 
6991
\\
$10^{10}$ 
& 
940
& 
2125
& 
3316
& 
4508
& 
5701
\\
$10^5$ 
& 
678
& 
1511
& 
2347
& 
3184
& 
4021
\\ 
\hline
\noalign{\kern 7pt}
\hline 
&\multicolumn{5}{c|}{$W^{(n)}$}
\\
\hline
$z_L$ $\backslash$ $n$ 
& 1 & 2 & 3 & 4 & 5
\\ \hline
$10^{15}$
& 
1133
& 
1800
& 
2587
& 
3285
& 
4050
\\
$10^{10}$
& 
927
& 
1470
& 
2111
& 
2679
& 
3301
\\
$10^5$
& 
659
& 
1041
& 
1490
& 
1889
& 
2325
\\
\hline
\noalign{\kern 7pt}
\hline 
&\multicolumn{5}{c|}{$Z^{(n)}$}
\\
\hline
$z_L$ $\backslash$ $n$ & 1 & 2 & 3 & 4 & 5
\\
\hline
$10^{15}$
& 
1130
& 
1803
& 
2584
& 
3289
& 
4046
\\
$10^{10}$
& 
923
& 
1474
& 
2107
& 
2683
& 
3297
\\
$10^5$
& 
653
& 
1047
& 
1484
& 
1895
& 
2319
\\
\hline
\end{tabular}
\begin{tabular}{|c|ccccc|}
\noalign{\kern -79pt}
\hline
&\multicolumn{5}{c|}{
${W'}^{(n)}$, ${Z'}^{(n)}$}
\\
\hline
$z_L$ $\backslash$ $n$ & 1 & 2 & 3 & 4 & 5
\\
\hline
$10^{15}$
& 
1122
& 
2576
& 
4039
& 
5503
& 
6968
\\
$10^{10}$
& 
914
& 
2097
& 
3287
& 
4479
& 
5672
\\
$10^5$
& 
640
& 
1469
& 
2303
& 
3139
& 
3974
\\
\hline
\noalign{\kern 7pt}
\hline
&\multicolumn{5}{c|}{
$A^{\hat{4}(n)}$}
\\
\hline
$z_L$ $\backslash$ $n$ & 1 & 2 & 3 & 4 & 5
\\
\hline
$10^{15}$
& 
1788
& 
3274
& 
4748
& 
6218
& 
7687
\\
$10^{10}$
& 
1456
& 
2665
& 
5061
& 
6257
& 
7451
\\
$10^5$
& 
1020
& 
1867
& 
2708
& 
3547
& 
4384
\\
\hline
\end{tabular}
\end{center}
\end{table}

\subsection{KK quarks and leptons}

The mass eigenvalue equations for quarks (\ref{FermionKK4}) and
(\ref{FermionKK5}) and similar equations for  leptons contain
the bulk mass parameters $c_q, c_\ell$ and the ratios 
$\tilde{\mu}^q/\mu_2^q,  \tilde{\mu}^\ell/\mu_3^\ell$,
which are  determined  such that
their running masses are given in Table~\ref{tab:input}.
For the light quarks and leptons with $c>\onehalf$
the bulk mass parameters shift to larger values for smaller $z_L$, 
whereas for the heavy quarks with $c<\onehalf$
they become smaller.  
 $c_q$,  $c_\ell$, $\tilde{\mu}^q/\mu_2^q$ and  $\tilde{\mu}^\ell/\mu_3^\ell$
 are tabulated in Table~\ref{tab:c-mu}.
We note that $\tilde{\mu}^\ell/\mu_3^\ell \ll 1$ as neutrino masses are very small.

\begin{table}[htb]
\begin{center}
\caption{$c$ and $\tilde{\mu}/\mu_2$ for quarks and leptons.}
\label{tab:c-mu}
\vskip 5pt
\begin{tabular}{|c|ccc|ccc|}
\hline 
&\multicolumn{3}{c|}{$c_q$} &\multicolumn{3}{c|}{$c_\ell$}
\\
\hline
$z_L$
& $(u,d)$ & $(c,s)$ & $(t,b)$
& $(\nu_e, e)$
& $(\nu_\mu, \mu)$
& $(\nu_\tau, \tau)$
\\ 
\hline
$10^{15}$
& 
0.843
& 
0.679
& 
0.432
& 
0.900
& 
0.736
& 
0.646
\\
$10^{10}$
& 
1.018
& 
0.770
& 
0.395
& 1.104
& 
0.856
& 
0.720
\\
$10^5$
& 
1.548
& 
1.049
& 
0.268
& 
1.721
& 
1.222
& 
0.948
\\
\hline
\noalign{\kern 7pt}
\hline 
&\multicolumn{3}{c|}{$\tilde{\mu}^q/\mu_2^q$} 
&\multicolumn{3}{c|} {$\mu_3^\ell/\tilde{\mu}^\ell$}
\\
\hline
$z_L$
& $(u,d)$ & $(c,s)$ & $(t,b)$
& $(\nu_e, e)$
& $(\nu_\mu, \mu)$
& $(\nu_\tau, \tau)$
\\ 
\hline
$10^{15}$
& 
2.283
& 
0.0889
& 
0.0173
& 
$4.87\times 10^8$
& 
$1.14\times 10^{10}$
& 
$3.47\times 10^{10}$
\\
$10^{10}$
& 
2.283
& 
0.0889
& 
0.0175
& 
$4.87\times 10^8$
& 
$1.14\times 10^{10}$
& 
$3.47\times 10^{10}$
\\
$10^5$
& 
2.283
& 
0.0889
& 
0.0181
& 
$4.87\times 10^8$
& 
$1.14\times 10^{10}$
& 
$3.47\times 10^{10}$
\\
\hline
\end{tabular}
\end{center}
\end{table}

The spectra of KK modes of quarks and leptons are determined by the 
same mass eigenvalue equations as the zero modes.  
They are tabulated  in Table~\ref{tab:mkkud}.
Except for the  KK tower of $(t,b)$, $u^{(n)}$ and $d^{(n)}$, for instance,
have approximately degenerate masses.
Similarly to the case of the gauge bosons,
we find an inequality
$m_{t^{(1)}} < m_{b^{(1)}}
< m_{c^{(1)}} \simeq m_{s^{(1)}} 
< m_{d^{(1)}} \simeq m_{u^{(1)}}$.

\begin{table}[htb]
\begin{center}
\caption{The masses of KK  quarks and leptons in unit of GeV.}
\label{tab:mkkud}
\vskip 5pt
\begin{tabular}{|c|ccc|} 
\hline
$z_L$ & $10^{15}$ & $10^{10}$ & $10^5$
\\
\hline 
$u^{(1)}, d^{(1)}$
& 
1361
& 
1203
& 
1037
\\
$u^{(2)}, d^{(2)}$
& 
2001
& 
1716
& 
1383
\\
$u^{(3)}, d^{(3)}$
& 
2823
& 
2397
& 
1886
\\
$u^{(4)}, d^{(4)}$
& 
3503
& 
2944
& 
2258
\\
$u^{(5)}, d^{(5)}$
& 
4287
& 
3590
& 
2727
\\
\hline
\end{tabular}
\hskip 3pt
\begin{tabular}{|c|ccc|} 
\hline
$z_L$ & $10^{15}$ & $10^{10}$ & $10^5$
\\
\hline 
$c^{(1)}, s^{(1)}$
& 
1249
& 
1068
& 
855
\\
$c^{(2)}, s^{(2)}$
& 
1900
& 
1593
& 
1213
\\
$c^{(3)}, s^{(3)}$
& 
2706
& 
2255
& 
1692
\\
$c^{(4)}, s^{(4)}$
& 
3394
& 
2812
& 
2075
\\
$c^{(5)}, s^{(5)}$
& 
4169
& 
3447
& 
2529
\\
\hline
\end{tabular}
\vskip 7pt
\begin{tabular}{|c|ccc|} 
\hline
$z_L$ & $10^{15}$ & $10^{10}$ & $10^5$
\\
\hline 
$t^{(1)}$
& 
1121
& 
912
& 
634
\\
$t^{(2)}$
& 
1797
& 
1467
& 
1037
\\
$t^{(3)}$
& 
2576
& 
2097
& 
1467
\\
$t^{(4)}$
& 
3279
& 
2672
& 
1877
\\
$t^{(5)}$
& 
4039
& 
3287
& 
2303
\\
\hline
\end{tabular}
\hskip 3pt
\begin{tabular}{|c|ccc|} 
\hline
$z_L$ & $10^{15}$ & $10^{10}$ & $10^5$
\\
\hline 
$b^{(1)}$
& 
1172
& 
975
& 
734
\\
$b^{(2)}$
& 
1745
& 
1402
& 
936
\\
$b^{(3)}$
& 
2627
& 
2160
& 
1567
\\
$b^{(4)}$
& 
3228
& 
2608
& 
1778
\\
$b^{(5)}$
& 
4090
& 
3351
& 
2402
\\
\hline
\end{tabular}
\vskip 7pt
\begin{tabular}{|c|ccc|} 
\hline
$z_L$ & $10^{15}$ & $10^{10}$ & $10^5$
\\
\hline 
$\nu_e^{(1)}, e^{(1)}, \nu_\mu^{(1)}, \mu^{(1)}, \nu_\tau^{(1)}, \tau^{(1)}$
& 
$1400$
& 
$1249$
& 
$1099$
\\
$\nu_e^{(2)}, e^{(2)}, \nu_\mu^{(2)}, \mu^{(2)}, \nu_\tau^{(2)}, \tau^{(2)}$
& 
$2036$
& 
$1758$
& 
$1441$
\\
$\nu_e^{(3)}, e^{(3)}, \nu_\mu^{(3)}, \mu^{(3)}, \nu_\tau^{(3)}, \tau^{(3)}$
& 
$2863$
& 
$2445$
& 
$1952$
\\
$\nu_e^{(4)}, e^{(4)}, \nu_\mu^{(4)}, \mu^{(4)}, \nu_\tau^{(4)}, \tau^{(4)}$
& 
$3540$
& 
$2990$
& 
$2321$
\\
$\nu_e^{(5)}, e^{(5)}, \nu_\mu^{(5)}, \mu^{(5)}, \nu_\tau^{(5)}, \tau^{(5)}$
& 
$4328$
& 
$3640$
& 
$2794$
\\
\hline
\end{tabular}
\end{center}
\end{table}

\section{Couplings  of KK gauge bosons \label{sec:fkkg} }

The couplings of quarks and leptons to KK gauge bosons 
can be calculated in the same manner as given in Sec.~4 for the 
couplings to the 4D gauge bosons.
As a general feature left-handed quarks and leptons are localized
near the Planck brane, whereas right-handed ones near the TeV brane.
KK gauge bosons are localized near the TeV brane so that right-handed
quarks and leptons have larger couplings than left-handed ones.
Because of this asymmetry the left-right symmetry is broken even in the 
strong interaction sector.

KK gluons do not decay into massless gluons.  On the other hand, 
KK $W$ and $Z$  can decay into $WZ$ and $WW$, respectively.

\bigskip

\leftline{\bf (i) KK photons and gluons}

The couplings of the first KK photon and gluon with 
quarks or leptons are tabulated in Table~\ref{kkphoton1coupling}.
The wave functions of the KK photon and gluon are  the same 
and their couplings to quarks are the same up to a normalization factor.
The largest coupling is $g_{u_R}^{G^{(1)}} \simeq g_{d_R}^{G^{(1)}}$.
This is different from other scenario in which
the $t$ quark  dominantly couples to  KK gluons.

\begin{table}[htb]
\begin{center}
\caption{The couplings of the first KK photon to leptons and quarks,
$g_f^{\gamma^{(1)}}/( g_A/\sqrt{L})$.  $e= (g_A/\sqrt{L}) \sin \theta_W$.
The couplings of the first KK gluon to  quarks, $g_f^{G^{(1)}}/( g_C/\sqrt{L})$,
are the same as $g_f^{\gamma^{(1)}}/( g_A/\sqrt{L})$.}
\label{kkphoton1coupling}
\vskip 5pt
\begin{tabular}{|c|ccc|ccc|}
\hline
$z_L$ 
& $e_L$ & $\mu_L$ & $\tau_L$
& $e_R$ & $\mu_R$ & $\tau_R$
\\
\hline
$10^{15}$
&
$-0.195$  
&
$-0.195$ 
& 
$-0.195$
& 
6.408
&
6.147
& 
5.981
\\
$10^{10}$
&
$-0.241$
& 
$-0.241$
& 
$-0.241$
&
5.426
& 
5.153
& 
4.968
\\
$10^5$
&
$-0.347$
&
$-0.347$
&
$-0.346$
&
4.123  
&
3.872
& 
3.672
\\ \hline
\noalign{\kern 5pt}
\hline
$z_L$ & $u_L$ & $c_L$ & $t_L$ & $u_R$ & $c_R$ & $t_R$
\\
\hline
$10^{15}$
&
$-0.195$ 
&
$-0.195$ 
&
$0.442$ 
&
$6.323$
& 
$6.044$ 
&
$5.603$ 
\\
$10^{10}$
&
$-0.241$ 
&
$-0.241$ 
&
$0.554$ 
&
$5.339$ 
&
$5.040$ 
&
$4.497$ 
\\
$10^5$
&
$-0.347$ 
&
$-0.347$ 
&
$0.890$ 
&
$4.049$ 
&
$3.753$ 
&
$2.925$ 
\\ 
\hline
\noalign{\kern 5pt}
\hline
$z_L$ & $d_L$ & $s_L$ & $b_L$
& $d_R$ & $s_R$ & $b_R$
\\
\hline
$10^{15}$
&
$-0.195$
&
$-0.195$
&
$0.661$ 
&
6.323 
&
6.044
&
5.500 
\\
$10^{10}$
&
$-0.241$
&
$-0.241$
&
$0.797$ 
&
5.339
&
5.040
&
4.376 
\\
$10^5$
&
$-0.347$
&
$-0.347$
&
$1.111$ 
&
4.049
&
3.753 
&
2.786
\\ 
\hline
\end{tabular}
\end{center}
\end{table}

We note that the couplings of right-handed fermions are
so large that the perturbative treatment  is not valid for the KK gluons.
With this reservation in mind one can evaluate the decay widths of the first KK gluon
by using the couplings  in Table~\ref{kkphoton1coupling}.  The decay width is given by
\beeq
\Gamma( G^{(n)} \to f\bar{f} )   
= C  \,  \frac{\alpha_s m_{G^{(n)}}}{ 6 } \, 
F(\bar g_{f L}^{G^{(n)}}, {\bar g}_{f R}^{G^{(n)}}, m_f, m_{G^{(n)}} ) ~, 
\label{KKgluondecay1}
\eneq
where $F$ is defined  in (\ref{Zdecay1}) and 
$\bar g_{f L}^{G^{(n)}} = g_{f L}^{G^{(n)}}/(g_C/\sqrt{L})$.
The color factor $C=3$. 
Numerical values are tabulated in Table~\ref{tab:kk1gdecay}.
It is found that the decay rate to the light quarks  is large.
The total decay width of $G^{(1)}$ turns out much larger than its mass.
Thus the KK gluon cannot be identified as a resonance.

\begin{table}[htb]
\begin{center}
\caption{First KK gluon decay:
the branching fraction and the total width at the tree level without QCD corrections.}
\label{tab:kk1gdecay}
\vskip 7pt
\renewcommand{\arraystretch}{1.1}
\begin{tabular}{|c|ccc|}
\hline
$z_L$ & $10^{15}$ & $10^{10}$ & $10^5$
\\
\hline
$u$ ($\%$)  &18.68 &19.40 & 20.88
\\
$d$ ($\%$) &18.68 &19.40 & 20.88
\\
$s$ ($\%$)  &17.07  &17.29  &17.96
\\
$c$ ($\%$)  &17.07  &17.29  &17.96
\\
$b$ ($\%$)  &14.33  &13.44  &11.38
\\
$t$ ($\%$)  &14.17  &13.19   &10.93
\\
\hline
$\Gamma_{\rm total}$ (GeV) &7205   &4070   &1576
\\
\hline
mass (GeV) & 1144 &940 & 678
\\
\hline
\end{tabular}
\end{center}
\end{table}

The decay width of the first KK photon $A^{\gamma (1)}$ is evaluated similarly.
The decay width to a fermion pair is
\beeq
\Gamma( \gamma^{(n)} \to f\bar{f} )   
= C~  \frac{q_f^2 \, \alpha \, m_{\gamma^{(n)}}}{3 \sin^2 \theta_W }  
~ F( \bar g_{f L}^{\gamma^{(n)}}, {\bar g}_{f R}^{\gamma^{(n)}}, 
m_f, m_{\gamma^{(n)}} ) ~, 
\eneq
where $\bar g_{f L}^{\gamma^{(n)}} =  g_{f L}^{\gamma^{(n)}} / ( g_A/\sqrt{L})$
and  $q_f$ is a charge $\frac{2}{3}, -\frac{1}{3}, -1, 0$.
$C =1$ for leptons.

In addition to decay into $q \bar q$ and $\ell \bar \ell$, the first KK photon can
decay into $W^+ W^-$ through
\beqn
&&\hskip -1.cm
{\cal L}_{\rm int}^{WW \gamma^{(n)}} 
= i g_{WW \gamma^{(n)}} \Big\{ 
(\dd_\mu W_\nu^\dagger - \dd_\nu W_\mu^\dagger ) W^\mu A^{\gamma (n) \nu}  \cr
\noalign{\kern 5pt}
&&\hskip 1.8cm
- (\dd_\mu W_\nu - \dd_\nu W_\mu ) W^{\dagger\mu} A^{\gamma (n) \nu}  
+ (\dd_\mu  A^{\gamma (n)}_\nu - \dd_\nu  A^{\gamma (n)}_\mu ) 
W^{\dagger\mu} W^\nu \Big\} ~,  \cr
\noalign{\kern 5pt}
&&\hskip -1.cm
g_{WW\gamma^{(n)}} = g_A \int  \frac{dz}{kz} \bigg[ 
h_{\gamma^{(n)}}^L \Big\{ ( h_{W}^L)^2 + \onehalf (h_{W}^\wedge)^2 \Big\} 
+ h_{\gamma^{(n)}}^R \Big\{ ( h_{W}^R)^2 + \onehalf (h_{W}^\wedge)^2 \Big\} 
\bigg] ~. 
\label{WWgamma1}
\eeqn
Inserting the mode functions in Appendix A, one finds
\beeq
g_{WW\gamma^{(n)}} = 
\frac{e \sqrt{L}}{\sqrt{ r_{\gamma^{(n)}} } \,  r_W}  \int \frac{dz}{kz}   ~
C(z; \lambda_{\gamma^{(n)}} ) \Big\{ C(z; \lambda_{W} )^2 
+ \hat S(z; \lambda_{W} )^2 \Big\} ~.
\label{WWgamma2}
\eneq
Note that the photon coupling is universal; 
$g_{WW\gamma} = g_{WW\gamma^{(0)}} = e$.
The first KK photon has a coupling 
$g_{WW\gamma^{(1)}} / e = (-0.05603, -0.06765,  -0.09145)$
for $z_L = (10^{15}, 10^{10}, 10^5)$.
The decay width is given by \cite{delAguila:1986ad}
\beeq
\Gamma(\gamma^{(n)} \to   W^+ W^-)   
=\frac{g_{WW \gamma^{(n)}}^2 m_{\gamma^{(n)}} }{192 \pi \eta_n^2} 
( 1+ 20 \eta_n + 12 \eta_n^2 ) (1 - 4 \eta_n)^{3/2}  
\label{KKgammaWWdecay}
\eneq
where $\eta_n = {m_W^2}/{m_{\gamma^{(n)}}^2}$.

The decay widths of the first KK photon are summarized in Table \ref{tab:kkph1decay}.  
The observed mass $m_W$ is used in the phase space of the final state 
in the evaluation of $\Gamma[\gamma^{(1)} \go W^+ W^-]$.
The first KK photon $A^{\gamma (1)}$ has a total decay width
larger than or comparable to its mass.   It does not look like a resonance.
\begin{table}[htb]
\begin{center}
\caption{Branching fractions and decay widths  of the first KK photon 
$\gamma^{(1)}$.   $\alpha = 1/128$ is used.}
\label{tab:kkph1decay}
\vskip 5pt
\renewcommand{\arraystretch}{1.1}
\begin{tabular}{|c|ccc|}
\hline
$z_L$ & $10^{15}$ & $10^{10}$ & $10^5$
\\
\hline
$e$ ($\%$) 
& 
13.5
& 
14.0
& 
14.8
\\
$\mu$ ($\%$)
& 
12.5
& 
12.6
& 
13.1
\\
$\tau$ ($\%$)
& 
11.8
& 
11.7
& 
11.8 
\\
\hline
$u$ ($\%$)
&
18.2
& 
18.8
& 
19.8
\\
$c$ ($\%$)
&
16.7
&
16.7
& 
17.0
\\
$t$ ($\%$)
& 
13.8
&
12.8
& 
10.4
\\
$d$ ($\%$)
&
4.56
& 
4.69
& 
4.95
\\
$s$ ($\%$)
& 
4.16
& 
4.18
& 
4.26
\\
$b$ ($\%$)
&
3.49
& 
3.25
& 
2.69
\\
\hline
$W$ ($\%$)
& 
1.30
&
1.28
& 
1.23
\\
\hline
$\Gamma[ ~{\rm all}~ f \bar f \,]$ (GeV)
& 
1933
&
1105
& 
441
\\
$\Gamma[W^+ W^-]$ (GeV)
& 
25.5
&
14.3
& 
5.5
\\
\hline
$\Gamma_{\rm total}$ (GeV)
& 
1959
& 
1120
&
446
\\
\hline
mass (GeV) & 1144 &940 & 678
\\
\hline
\end{tabular}
\end{center}
\end{table}

\leftline{\bf (ii) KK $W$ and $Z$}

The coupling of quarks and leptons to the first KK $W$ boson
are given in Table~\ref{tab:kkw1q}.
The quarks in the third generation
have larger couplings than the other quarks and leptons.
Couplings of right-handed quarks and leptons are rather small.

The fermion couplings of the first KK $Z$ boson can be calculated similarly.
They are tabulated in Table~\ref{kkz1coupling}.   The values of the couplings 
of left-handed leptons are not very sensitive to the generation.
For a smaller warp factor, the magnitude of the couplings of left-handed
(right-handed)  leptons and quarks become larger (smaller).
For left-handed leptons and quarks, 
the third generation has larger couplings than the first and second generations.
In contrast, for right-handed leptons and quarks,
the third generation has smaller couplings.

\begin{table}[htb]
\begin{center}
\caption{The couplings of the first KK $W$ boson
with quarks and leptons, 
$g_f^{W^{(1)}} \sqrt{L} /g_A$.}
\label{tab:kkw1q}
\vskip 5pt
\begin{tabular}{|c|ccc|}
\hline
$z_L$  & $u_L d_L$ & $c_L s_L$ & $t_L b_L$
\\ \hline
$10^{15}$
&
$-0.138$
& $-0.138$ 
& 
0.492
\\
$10^{10}$
& 
$-0.170$
& 
$-0.170$
& 
0.609
\\
$10^{5}$
& 
$-0.244$
& 
$-0.244$
& 
0.934
\\ 
\hline
\noalign{\kern 3pt}
\hline
$z_L$
& $e_L \nu_{eL}$ & $\mu_L \nu_{\mu L}$ & $\tau_L \nu_{\tau L}$
\\
\hline
$10^{15}$
& 
$-0.138$
& 
$-0.138$
& 
$-0.138$
\\
$10^{10}$
& 
$-0.170$
& 
$-0.170$
& 
$-0.170$
\\
$10^5$
& 
$-0.244$
& 
$-0.244$
& 
$-0.244$
\\
\hline
\noalign{\kern 3pt}
\hline
$z_L$  & $u_R d_R$ & $c_R s_R$ & $t_R b_R$
\\ \hline
$10^{15}$
& 
$1.02 \times 10^{-12}$
&
$1.08 \times 10^{-8}$
& 
0.000308
\\
$10^{10}$
& 
$1.69 \times 10^{-12}$
& 
$1.88 \times 10^{-8}$
& 
0.000596
\\
$10^{5}$
& 
$3.66 \times 10^{-12}$
& 
$4.55 \times 10^{-8}$
& 
0.00204
\\ \hline
\end{tabular}
\end{center}
\end{table}
\begin{table}[htb]
\begin{center}
\caption{The couplings of the first KK $Z$ 
boson to leptons and quarks,
$g_f^{Z^{(1)}} \sqrt{L}/g_A$.}
\label{kkz1coupling}
\vskip 5pt
\begin{tabular}{|c|ccc|ccc|}
\hline
$z_L$
& $\nu_{eL}$ & $\nu_{\mu L}$ & $\nu_{\tau L}$
& $\nu_{eR}$ & $\nu_{\mu R}$ & $\nu_{\tau R}$
\\
\hline
$10^{15}$
& 
$-0.0577$
& 
$-0.0577$
& 
$-0.0576$
& {$1.1\times 10^{-31}$} 
& {$1.0\times 10^{-29}$} 
& {$3.5\times 10^{-28}$} 
\\
$10^{10}$
& 
$-0.0712$
& 
$-0.0712$
& 
$-0.0711$
& {$1.8\times 10^{-31}$} 
& {$1.7\times 10^{-29}$} 
&{$6.1\times 10^{-28}$} 
\\
$10^5$
& 
$-0.1025$
& 
$-0.1025$
& 
$-0.1023$
& {$3.8\times 10^{-31}$} 
& {$4.0\times 10^{-29}$} 
& {$1.5\times10^{-27}$} 
\\
\hline
\noalign{\kern 5pt}
\hline
$z_L$ 
& $e_L$ & $\mu_L$ & $\tau_L$
& $e_R$ & $\mu_R$ & $\tau_R$
\\
\hline
$10^{15}$
& 
0.0311
& 
0.0310
& 
0.0311
& 
2.516
& 
2.420
& 
2.352
\\
$10^{10}$
& 
0.0384
& 
0.0383
& 
0.0384
& 
2.126
& 
2.033
& 
1.953
\\
$10^5$
& 
0.0557
& 
0.0553
& 
0.0558
& 
1.598
& 
1.531
& 
1.436
\\ \hline
\noalign{\kern 5pt}
\hline
$z_L$ & $u_L$ & $c_L$ & $t_L$ & $u_R$ & $c_R$ & $t_R$
\\
\hline
$10^{15}$
& 
$-0.0400$
& 
$-0.0400$
& 
$-0.2058$
& 
$-1.656$
& 
$-1.585$
& 
$-1.467$
\\
$10^{10}$
& 
$-0.0493$
& 
$-0.0493$
& 
$-0.2553$
& 
$-1.396$
& 
$-1.320$
& 
$-1.174$
\\
$10^5$
& 
$-0.0713$
& 
$-0.0713$
& 
$-0.3814$
& 
$-1.048$
& 
$-0.977$
& 
$-0.743$
\\ 
\hline
\noalign{\kern 5pt}
\hline
$z_L$ & $d_L$ & $s_L$ & $b_L$
& $d_R$ & $s_R$ & $b_R$
\\
\hline
$10^{15}$
& 
0.0488
& 
0.0488
& 
$-0.5581$
& 
0.828
& 
0.792
& 
0.723
\\
$10^{10}$
& 
0.0602
& 
0.0602
& 
$-0.6710$
& 
0.698
& 
0.660
& 
0.576
\\
$10^5$
& 
0.0869
& 
0.0869
& 
$-0.9219$
& 
0.524
& 
0.488
& 
0.371
\\ 
\hline
\end{tabular}
\end{center}
\end{table}

Just like KK photons KK $Z$ bosons can decay into a pair of $W$ bosons.  
Their couplings  are given by 
\beqn
&&\hskip -1.cm
{\cal L}_{\rm int}^{WWZ^{(n)}} 
= i g_{WWZ^{(n)}} \Big\{ 
(\dd_\mu W_\nu^\dagger - \dd_\nu W_\mu^\dagger ) W^\mu Z^{(n) \nu}  \cr
\noalign{\kern 5pt}
&&\hskip 1.5cm
- (\dd_\mu W_\nu - \dd_\nu W_\mu ) W^{\dagger\mu} Z^{(n) \nu}  
+ (\dd_\mu  Z^{(n)}_\nu - \dd_\nu  Z^{(n)}_\mu ) W^{\dagger\mu} W^\nu \Big\} ~,  \cr
\noalign{\kern 10pt}
&&\hskip -1.cm
g_{WWZ^{(n)}} = g_A \int  \frac{dz}{kz} \bigg[ 
h_{Z^{(n)}}^L \Big\{ ( h_{W}^L)^2 + \onehalf (h_{W}^\wedge)^2 \Big\} 
+ h_{Z^{(n)}}^R \Big\{ ( h_{W}^R)^2 + \onehalf (h_{W}^\wedge)^2 \Big\} \cr
\noalign{\kern 5pt}
&&\hskip 7.5cm
+ h_{Z^{(n)}}^\wedge ( h_{W}^L + h_{W}^R ) h_{W}^\wedge \bigg] ~, 
\label{WWZint1}
\eeqn
where  indices $\mu,\nu$ are contracted with $\eta_{\mu\nu}$.
With mode functions inserted, 
\beqn
&&\hskip -1.cm
\frac{g_{WWZ^{(n)}}}{~ \myfrac{g_A}{\sqrt{L}} \cos\theta_W ~} \equiv I_{WWZ^{(n)}} \cr
\noalign{\kern 10pt}
&&\hskip -1.cm
= \frac{\sqrt{L}}{\sqrt{ 2 r_{Z^{(n)}} } \,  r_W}  \int \frac{dz}{kz}  
\bigg[ \frac{1 - 2 \sin^2 \theta_W}{\cos^2 \theta_W}
 C(z; \lambda_{Z^{(n)}} ) \Big\{ C(z; \lambda_{W} )^2 + \hat S(z; \lambda_{W} )^2 \Big\} \cr
\noalign{\kern 10pt}
&&\hskip 5.0cm
+  \frac{2}{\cos^2 \theta_W} \hat S(z; \lambda_{Z^{(n)}} )  C(z; \lambda_{W} )
\hat S(z; \lambda_{W} ) \bigg] ~,   
\label{WWZint2}
\eeqn
With the couplings $g_{WWZ^{(n)}}$ the partial decay width 
$\Gamma(Z^{(n)} \to   W^+ W^-)$  is given by the formula (\ref{KKgammaWWdecay})
where $g_{WW\gamma^{(n)}}$ and  $m_{\gamma^{(n)}}$ are replaced by
$g_{WWZ^{(n)}}$ and  $m_{Z^{(n)}}$, respectively.
The enhancement factor $1/\eta_n^2 = (m_{Z^{(n)}}/m_W)^4$ represents
that $Z^{(n)}$ decays dominantly to the  longitudinal components of $W$
over the transverse components.

The numerical values of the couplings $g_{WWZ^{(n)}}$ are tabulated 
in Table \ref{WWZcoupling}.   $g_{WWZ} \equiv g_{WWZ^{(0)}}$ has been 
evaluated in \cite{HS2}. There appears tiny deviation in $g_{WWZ}$ from
that in the SM.  The couplings of KK $Z$ are found very small; 
$ |g_{WWZ^{(n)}}| \ll g_{WWZ}$.

\begin{table}[htb]
\begin{center}
\caption{The couplings $WWZ^{(n)}$.  
The ratios $I_{WWZ^{(n)}} = g_{WWZ^{(n)}}/(g_A\cos \theta_W/\sqrt{L})$ are listed.
$n=0$ corresponds to the $WWZ$ coupling.}
\label{WWZcoupling}
\vskip 5pt
\begin{tabular}{|c|c|c|c|}
\hline
$z_L$ & $10^{15}$ & $10^{10}$ & $10^5$ 
\\
\hline
$WWZ$ & 0.99985 & 0.99966 & 0.99862 
\\
\hline
$WWZ^{(1)}$ & -0.0343 & -0.0422 &  -0.0604
\\
\hline
$WWZ^{(2)}$ &  $2.07 \times 10^{-5}$ &  $3.35 \times 10^{-5}$ &  $5.42 \times 10^{-5}$
\\
\hline
$WWZ^{(3)}$ & $- 1.25 \times 10^{-3}$ & $- 1.55 \times 10^{-3}$  &  $- 2.26 \times 10^{-3}$
\\
\hline
$WWZ^{(4)}$ & $- 1.38 \times 10^{-5}$  & $- 2.59 \times 10^{-5}$  &  $- 7.76 \times 10^{-5}$
\\
\hline
$WWZ^{(5)}$ & $- 2.04 \times 10^{-4}$ & $- 2.50 \times 10^{-4}$ &  $- 3.56 \times 10^{-4}$
\\
\hline
\end{tabular}
\end{center}
\end{table}

The decay width of the first KK $Z$ boson is tabulated in Table~\ref{tab:kk1zdecay}.
The mass and total decay width of $Z^{(1)}$ are   1130$\,$GeV and  422$\,$GeV
for $z_L = 10^{15}$,  respectively.
The branching fraction of the $WW$ mode is about 7\%.
(The observed mass $m_W$ is used in the phase space of the final state 
in the evaluation of $\Gamma[Z^{(1)} \go W^+ W^-]$.)
In contrast to the decay width of $Z$ boson given in Table~\ref{tab:decay},
the decay rates for neutrinos in  the first KK $Z$ boson decay are very small.  

\begin{table}[htb]
\begin{center}
\caption{First KK $Z$ boson decay:
the branching fractions and decay widths. $\alpha=1/128$ is used.}
\label{tab:kk1zdecay}
\vskip 5pt
\renewcommand{\arraystretch}{1.1}
\begin{tabular}{|c|ccc|}
\hline
$z_L$ & $10^{15}$ & $10^{10}$ & $10^5$
\\
\hline
$e$ ($\%$) 
& 
12.4
& 
12.5
& 
11.8
\\
$\mu$ ($\%$)
& 
11.5
& 
11.4
& 
10.9
\\
$\tau$ ($\%$)
& 
10.9
& 
10.6
& 
9.56
\\
$\nu_e + \nu_\mu + \nu_\tau$ ($\%$)
& 
0.02
& 
0.04
& 
0.15
\\
\hline
$u$ ($\%$)
&
16.7
& 
16.8
& 
15.9
\\
$c$ ($\%$)
&
15.3
&
15.0
& 
13.8
\\
$t$ ($\%$)
& 
12.9
&
11.9
& 
9.51
\\
$d$ ($\%$)
&
4.20
& 
4.23
& 
4.06
\\
$s$ ($\%$)
& 
3.85
& 
3.79
& 
3.55
\\
$b$ ($\%$)
&
5.09
& 
6.74
& 
14.2
\\
\hline
$W$ ($\%$)
& 
7.10
&
6.96
& 
6.51
\\
\hline
$\Gamma [W^+ W^- ]$ (GeV)
& 
30.0
&
17.0
& 
6.8
\\
\hline
$\Gamma_{\rm total}$ (GeV)
& 
422
& 
245
&
104
\\
\hline
mass (GeV) & 1130 & 923 & 653
\\ \hline
\end{tabular}
\end{center}
\end{table}

\section{Signals of KK $Z$  at Tevatron and LHC}

\begin{figure}[htb]
\begin{center}
\includegraphics[width=5cm]{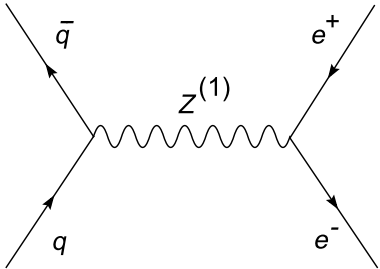}
\caption{The first KK $Z$ boson signal.}
\label{kkzsignal}
\end{center}
\end{figure}

The KK $Z$ boson can be produced at Tevatron and LHC.
We first consider the production process of the first KK $Z$ boson ($Z^{(1)}$)
followed by its decay into an electron and a positron, 
$q \bar q \go Z^{(1)} \go e^+ e^-$, as  shown in Fig.~\ref{kkzsignal}.
To this process the first KK photon ($A^{\gamma(1)}$) also contributes,
which has a mass close to that of $Z^{(1)}$.  
Unlike $Z^{(1)}$, however, $A^{\gamma (1)}$ has a decay width larger than 
its mass so that its contribution is expected to give an additional
smooth background to the $Z^{(1)}$ signal. 
Effects of KK particles such as $A^{\gamma(n)}$ ($n\geq 2$) are ignored 
in our analysis for simplicity, though they also give smooth background.
Our numerical calculation is done by MadGraph/MadEvent \cite{MGMEV4} 
at the parton level with CTEQ6L parton distribution function \cite{CTEQ6L}
and without detector simulation.

The cross sections of $p\bar p\to e^+e^-X$ at
$\sqrt{s}=1.96\ \mathrm{TeV}$ are evaluated as 22, 7.1 and 3.8 pb for
$z_L=10^5$, $10^{10}$ and $10^{15}$,  respectively, where
the invariant mass of the charged leptons is required to be
larger than 150 GeV, and other cuts are the default values of
MadGraph/MadEvent: $p_T>10\,\mathrm{GeV}$, $|\eta|<2.5$, and $\Delta R>0.4$
for the charged leptons. In the current model the production rate of 
$Z^{(1)}$ decreases for larger $z_L$ as it becomes heavier.  
The background cross section, that is, the Drell-Yan cross section in 
the SM is 0.73 pb. Including 10\% theoretical uncertainty in the signal 
estimation, we obtain the statistical significance at Tevatron with 
the integrated luminosity of $5.4 \, (2.5)\ \mathrm{fb}^{-1}$, which
corresponds to the analysis by D0 collaboration \cite{D0ZP2011} 
(CDF collaboration \cite{CDFZP2009}),  as $9.7 \, (9.7), 9.0 \, (8.9)$ and 
$8.1 \, (8.0)$  for $z_L=10^5$, $10^{10}$ and $10^{15}$,  respectively.

\begin{figure}[htb]
\begin{center}
\vskip .5cm
\includegraphics[width=13.cm]{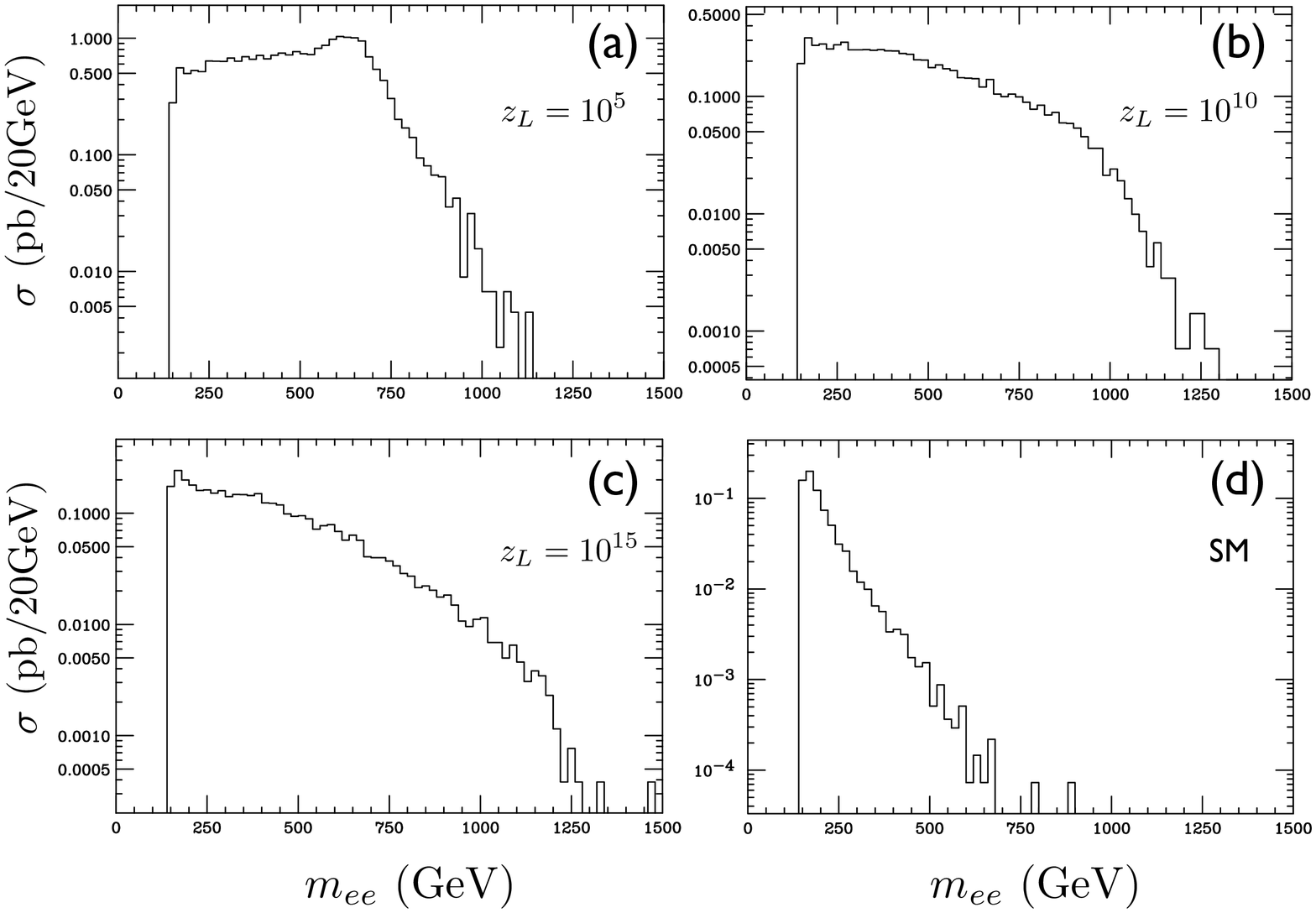}
\caption{Distributions of the $e^+e^-$ invariant mass
in  $p\bar p\to e^+e^-X$ at $\sqrt{s}=1.96\ \mathrm{TeV}$.
(a) The present model with $z_L=10^5$.
(b) $z_L=10^{10}$.
(c) $z_L=10^{15}$.
(d) The SM.}
\label{TevatronKZKA}
\end{center}
\end{figure}

The first KK $Z$ corresponds to what is referred to as $Z'$
in the analysis of Tevatron data \cite{D0ZP2011,CDFZP2009}.
So far no signal of $Z'$ has been found, which gives a constraint 
on the present model.  
The signals expected at Tevatron are depicted in 
Fig.~\ref{TevatronKZKA}. A peak structure due to the first KK Z boson 
is seen in the case of $z_L=10^5$, and thus the scenario with 
$z_L = 10^5$ is excluded. Furthermore, although the KK Z resonance shape
is smeared out by the broad contribution of the first KK photon, 
the other scenarios with $z_L=10^{10}$ and $10^{15}$ also seem disfavored 
by the present Tevatron data based only on the total cross section 
as stated above. If we take the detailed invariant mass distribution
of the lepton pair and/or the dimuon channel into account, 
the limit on the warp factor will be strengthened.

\begin{figure}[htb]
\begin{center}
\vskip .5cm
\includegraphics[width=13.cm]{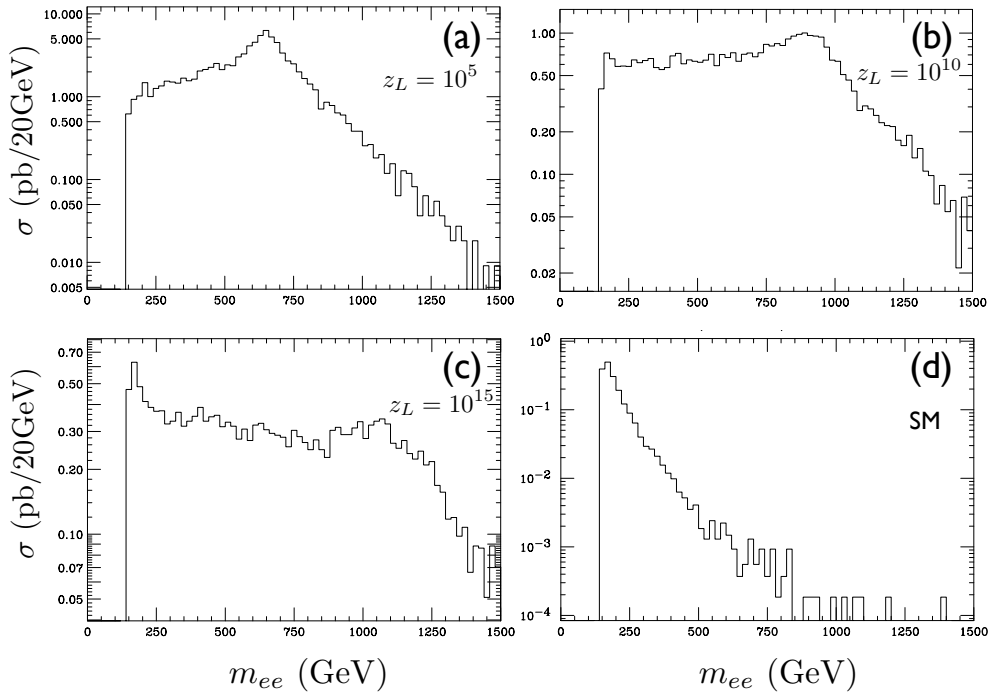}
\caption{Distributions of the $e^+e^-$ invariant mass in
$pp\to e^+e^-X$ at $\sqrt{s}=7\ \mathrm{TeV}$.
(a) The present model with $z_L=10^5$.
(b) $z_L=10^{10}$.
(c) $z_L=10^{15}$.
(d) The SM.}
\label{LHC7KZKA}
\end{center}
\end{figure}

As for LHC, we obtain the cross sections of $pp\to e^+e^-X$ at
$\sqrt{s}=7\ \mathrm{TeV}$ as 91, 36 and 20 pb for
$z_L=10^5$, $10^{10}$ and $10^{15}$ respectively, and 1.8 pb for the SM.
The same cuts on the final state as the $p\bar p$ case are applied.
The statistical significance at LHC is summarized in Table \ref{tab:LHCEE},
where 10\% theoretical uncertainty is assumed. 
The signals expected at LHC are shown in 
Fig.~\ref{LHC7KZKA}. The resonant structure of the first KK Z boson remains
for all the three values of $z_L$. 

Recently, CMS and ATLAS collaborations have searched for narrow 
resonances in dilepton channels and found no significant deviation from 
the SM \cite{CMSZP2011,ATLASZP2011}. The integrated luminosity for 
the electron channel is reported as $35\ \mathrm{pb^{-1}}$ by CMS
and $39\ \mathrm{pb^{-1}}$ by ATLAS.
Accordingly, the cases that $z_L \siml 10^{15}$ seems unlikely although
we need a detailed analysis to determine the excluded parameter region. 
It should be noted that the total decay width of the first KK $Z$
is very large in the current model, whereas a narrow width 
(3 \% of its mass or less) has been assumed in the analysis in 
Refs.\ \cite{CMSZP2011,ATLASZP2011}.

We comment that contributions from higher KK photons $A^{\gamma (n)}$ 
($n \ge 2$), which have broad decay widths,  may have destructive interference 
with that  from the first KK photon 
so that the magnitude of the smooth background is significantly decreased.
If this is the case, the bound from the current data at Tevatron and at LHC 
is weakened.  More thorough study is necessary on this respect, which is reserved 
for future.

\begin{table}[htb]
\begin{center}
\caption{Significance of $pp\to e^+e^-X$ at $\sqrt{s}=7\ \mathrm{TeV}$.}
\label{tab:LHCEE}
\vskip 5pt
\begin{tabular}{|c|ccc|}
\hline
$z_L$ & $10^{5}$ & $10^{10}$ & $10^{15}$ \\
\hline
$L=35\ \mathrm{pb}^{-1}$ & 9.7 & 9.1 & 8.5 \\
$L=100\ \mathrm{pb}^{-1}$ & 9.7 & 9.4 & 8.9 \\
$L=1000\ \mathrm{pb}^{-1}$ & 9.8 & 9.5 & 9.1 \\
\hline
\end{tabular}
\end{center}
\end{table}

\begin{figure}[htb]
\begin{center}
\vskip .7cm
\includegraphics[width=13cm]{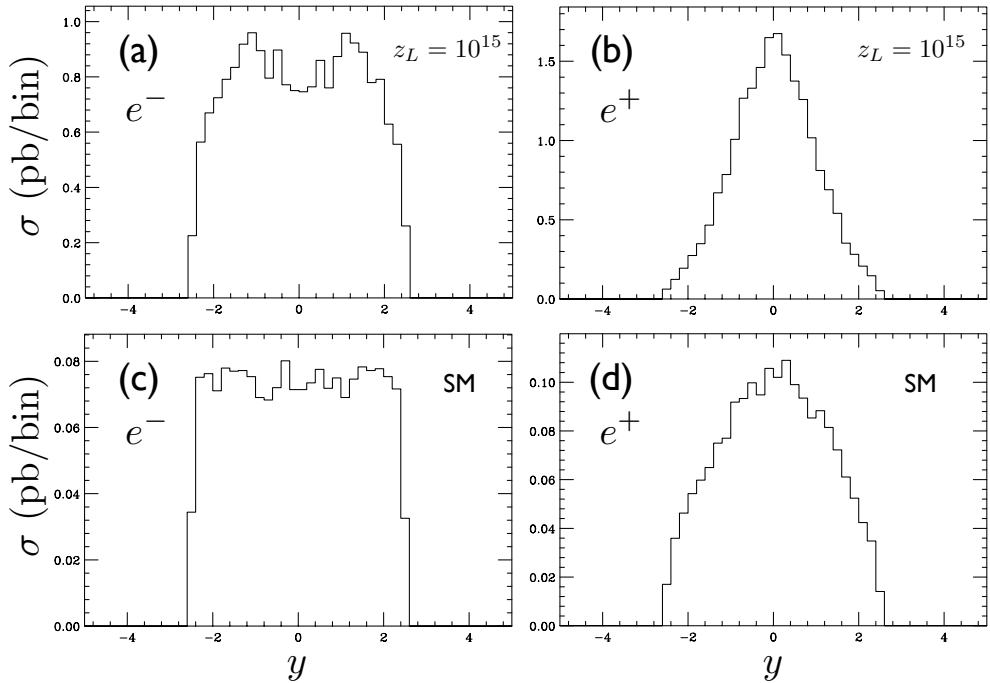}
\caption{The rapidity distributions. (a) The electron distribution
in the present model with $z_L=10^{15}$. (b) The positron distribution
in the present model with $z_L=10^{15}$. (c) The electron distribution
in the SM. (d) The positron distribution in the SM.}
\label{FIG:RPD}
\end{center}
\end{figure}

As seen in Tables \ref{kkphoton1coupling} and \ref{kkz1coupling}, 
the couplings of $A^{\gamma (1)}$ and $Z^{(1)}$
to the right-handed fermions except the neutrinos and the bottom
quark are significantly larger than those to the left-handed fermions.
Such a parity violation affects the distribution of the leptons in the final
state. Consider a favored parton-parton collision, for instance,  
$u_R\bar u_R\to e^-_R e^+_R$. The direction of the final $e^-_R$ tends
to be that of the initial $u_R$ because of the helicity conservation.
This angular distribution in the parton center-of-mass frame results 
in a harder electron spectrum (and a softer positron spectrum)
in the $pp$ center-of-mass frame since most of the initial quark-antiquark
pairs are boosted in the direction of the initial quark in 
the $pp$ collider. Hence, we expect a wider rapidity distribution for 
the electron than the positron. We present, in Fig.~\ref{FIG:RPD},
the rapidity ($y$) distributions of the electron and positron in
the present model with $z_L=10^{15}$ and in the SM. Though the both models
have the similar tendency that the electron distribution is
wider than the positron, the difference between the electron and 
the positron is more significant in the present model. This feature
in the rapidity distributions is quantified by the central charge 
asymmetry \cite{AKR},
\begin{equation}
A_{cc}(y_c)=\frac{\sigma(|y_{e^-}|<y_c)-\sigma(|y_{e^+}|<y_c)}
                 {\sigma(|y_{e^-}|<y_c)+\sigma(|y_{e^+}|<y_c)} \,  .
\end{equation}
Our numerical study suggests that the statistical significance of 
$A_{cc}(y_c)$ is maximized with $y_c\sim 0.6$ for the case of $z_L=10^{15}$.
We find that $A_{cc}(0.6)=-0.32(-0.17)$ for $z_L=10^{15}$ (the SM)
and the significance of $5\sigma$ is expected with 
the integrated luminosity of about $1\,\mathrm{fb}^{-1}$. 
Another signal of the parity violation may be seen
in the lepton forward-backward asymmetry with respect to the
boost direction of the KK $Z$ boson \cite{Dittmar,DND}.

We also evaluate the cross section of $pp\to je^+e^-X$ at
$\sqrt{s}=7\ \mathrm{TeV}$, where $j$ denotes a jet,
to find 43, 17 and 9.3 pb for $z_L=10^5$, $10^{10}$ and $10^{15}$ 
respectively. The same cuts on the final leptons as the $pp\to e^+e^-X$ case
are applied, and the default cuts for jets in MadGraph/MadEvent, that is, 
$p_T>20\,\mathrm{GeV}$, $|\eta|<5$ and $\Delta R>0.4$ for the jet are employed.
Approximately 80\% of the cross section for $z_L=10^{15}$ includes
a gluon jet ($q\bar q\to gZ^{(1)}, g A^{\gamma (1)}$) and the rest does a quark jet 
($qg\to qZ^{(1)}, q A^{\gamma (1)}$). The SM cross section is estimated 
to be 0.66 pb. The statistical significance assuming 10\%
theoretical uncertainty is shown in Table \ref{tab:LHCJEE}.
The $pp\to je^+e^-X$ mode has a comparable sensitivity of 
the $pp\to e^+e^-X$ mode from the statistical point of view, 
but the background should be studied carefully 
since the signal is more complicated.

\begin{table}[htb]
\begin{center}
\caption{Significance of $pp\to je^+e^-X$ at $\sqrt{s}=7\ \mathrm{TeV}$.}
\label{tab:LHCJEE}
\vskip 5pt
\begin{tabular}{|c|ccc|}
\hline
$z_L$ & $10^{5}$ & $10^{10}$ & $10^{15}$ \\
\hline
$L=35\ \mathrm{pb}^{-1}$ & 9.5 & 8.9 & 8.1 \\
$L=100\ \mathrm{pb}^{-1}$ & 9.7 & 9.3 & 8.8 \\
$L=1000\ \mathrm{pb}^{-1}$ & 9.8 & 9.6 & 9.2 \\
\hline
\end{tabular}
\end{center}
\end{table}

\section{Conclusions}

In the present paper we have explored collider signatures of the 
$SO(5) \times U(1)$ gauge-Higgs unification model in the RS space.
The model predicts $\theta_H = \onehalf \pi$ and the stable Higgs boson.
The gauge and Higgs couplings of quarks and leptons deviate from those
in the standard model. 
With the warp factor $z_L$ given, the mass spectra and couplings
of all fields are determined.

There arises small deviation in the gauge couplings of quarks and leptons.
They lead to forward-backward asymmetry in $e^+ e^-$ annihilation
on the $Z$ resonance.  It was found that the gauge-Higgs unification model 
gives good fit to the 
forward-backward asymmetry
data 
in a wide rage of $z_L$.  However, the data of
branching fractions of various modes in the $Z$ boson decay is fit well 
only for $z_L \simg 10^{15}$.

Pair production of Higgs bosons at ILC, $e^+ e^- \go ZHH$, is marginal.
There is large background containing neutrinos. 
With polarized beam and appropriate cut, the statistical significance $S$ of 
the signal is estimated to be $S/\sqrt{L ({\rm fb}^{-1})} = 0.11$ 
for $\sqrt{s} = 750\,$GeV for $z_L = 10^{15}$, which requires 
the luminosity $L > 2.0\,$ab$^{-1}$ for $5 \sigma$ discovery.

Another important way to test the model is to produce KK modes.
The production of the first KK $Z$ boson $Z^{(1)}$ decaying into $e^+ e^-$ 
gives a clear signal.  At $z_L= 10^{15}$ the mass and width of $Z^{(1)}$ are
about 1130$\,$GeV and 422$\,$GeV, respectively.
The production of $Z^{(1)}$ can be discovered at LHC through
$pp \go Z^{(1)} X \go e^+ e^- X$ with 100$\,$pb$^{-1}$.
There appears a smooth background coming from the production and decay of 
KK photons.  The mass and width of the first KK photon $A^{\gamma (1)}$
are 1144$\,$GeV and 1959$\,$GeV  at $z_L= 10^{15}$, respectively.
We have evaluated the cross section including the contribution from $A^{\gamma (1)}$.
The present limit from the $Z'$ searches at Tevatron and LHC excludes
$z_L \siml 10^{15}$.  However, a more thorough study taking  account of
contributions of higher KK photons $A^{\gamma (n)}$ ($n \ge 2$) is necessary,
as destructive interference could occur in the smooth background.

It is a general feature that 
KK gluons, photons and $Z$  couple dominantly to right-handed quarks and leptons. 
The large parity violation  affects the rapidity distributions of
$e^+$ and $e^-$ in the decay of $Z^{(1)}$, which is quantified by measuring
the central charge asymmetry. 

We conclude that the present precision data of the gauge couplings 
and $Z'$ search indicates a large warp factor 
$z_L > 10^{15}$.
$Z^{(1)}$ production at LHC is a
promising way to test the model.

In the present paper we have investigated the $SO(5) \times U(1)$ gauge-Higgs
model with bulk fermions in the vector representation of $SO(5)$, in which right-handed
quarks and leptons are localized near the TeV brane and have large 
couplings with KK gauge bosons. 
It is interesting to see whether the couplings of the leptons to the KK photons
can be suppressed by introducing bulk lepton multiplets  in other
tensorial representation of $SO(5)$.

It has been shown that in order for the stable Higgs boson to account 
for the dark matter of the universe, its mass must be $m_H \sim 70\,$GeV,
which is obtained with a small warp factor $z_L \sim 10^5$ in the current model.
Further improvement of the model is necessary to explain both collider 
data and dark matter. 

\vskip 1cm

\leftline{\bf Acknowledgements}

The authors would like to thank C.\ Csaki, N.\ Haba, S.\ Kanemura and 
S.\ Matsumoto  for many valuable comments.
This work was supported in part 
by  Scientific Grants from the Ministry of Education and Science, 
Grant No.\ 20244028 (Y.H., N.U.),  
Grant No.\ 21244036 (Y.H.), and Grant No.\ 20244037 (M.T.).

\vskip 1cm

\begin{appendix}
\section{Normalized mode functions}

In this appendix, mode functions with their normalization factors
at $\theta_H = \onehalf \pi$  are collected.\cite{HNU}
Basis functions are given in (\ref{BesselF1}) and (\ref{Bessel2}).
For convenience we define
\beeq
\hat S(z; \lambda) = \frac{C(1; \lambda)}{S(1; \lambda)} \, S(z; \lambda) 
~,~~~
\hat S_L(z; \lambda, c ) = \frac{C_L(1; \lambda, c)}{S_L(1; \lambda, c)} 
\, S_L(z; \lambda, c) 
~.
\label{hatS}
\eneq

\subsubsection*{Gauge bosons}

Gauge bosons are expanded as in (\ref{expansion1}).  Mode functions $h(z)$
of $P_H$-even fields, for instance, satisfy orthogonality conditions
\beqn
&&\hskip -1.cm
\int_1^{z_L} \frac{dz}{kz} \Big\{ h_{W^{(n)}}^L h_{W^{(m)}}^L
+ h_{W^{(n)}}^R h_{W^{(m)}}^R 
+ h_{W^{(n)}}^\wedge h_{W^{(m)}}^\wedge \Big\} = \delta^{nm} ~, \cr
\noalign{\kern 10pt}
&&\hskip -1.cm
\int_1^{z_L} \frac{dz}{kz} \Big\{ h_{Z^{(n)}}^L h_{Z^{(m)}}^L
+ h_{Z^{(n)}}^R h_{Z^{(m)}}^R 
+ h_{Z^{(n)}}^\wedge h_{Z^{(m)}}^\wedge 
+ h_{Z^{(n)}}^B h_{Z^{(m)}}^B \Big\} = \delta^{nm} ~, \cr
\noalign{\kern 10pt}
&&\hskip -1.cm
\int_1^{z_L} \frac{dz}{kz} \Big\{ h_{\gamma^{(n)}}^L h_{\gamma^{(m)}}^L
+ h_{\gamma^{(n)}}^R h_{\gamma^{(m)}}^R 
+ h_{\gamma^{(n)}}^B h_{\gamma^{(m)}}^B \Big\} = \delta^{nm} ~, \cr
\noalign{\kern 10pt}
&&\hskip -1.cm
\int_1^{z_L} \frac{dz}{kz} \Big\{ h_{Z^{(n)}}^L h_{\gamma^{(m)}}^L
+ h_{Z^{(n)}}^R h_{\gamma^{(m)}}^R 
+ h_{Z^{(n)}}^B h_{\gamma^{(m)}}^B \Big\} = 0 ~.
\label{orthogonal1}
\eeqn
Similar relations hold for other mode functions.

\bigskip
\noindent
{\bf (i) Photon  tower ($\hat A_\mu^\gamma$)}
\beqn
&&\hskip -1.cm
h_{\gamma^{(n)}}^L = h_{\gamma^{(n)}}^R = 
 \frac{s_\phi}{\sqrt{1 + s_\phi^2}} 
\frac{1}{\sqrt{\strut  r_{\gamma^{(n)}}}}
C(z; \lambda_{\gamma^{(n)}} ) ~, \cr
\noalign{\kern 5pt}
&&\hskip -1.cm
h_{\gamma^{(n)}}^B  %\hskip 1.4cm 
= \frac{c_\phi}{\sqrt{1 + s_\phi^2}} 
\frac{1}{\sqrt{\strut  r_{\gamma^{(n)}}}}
C(z; \lambda_{\gamma^{(n)}} ) ~, \cr
\noalign{\kern 5pt}
&&\hskip -1.cm
r_{\gamma^{(n)}} = \int_1^{z_L} \frac{dz}{k z}\,  C(z; \lambda_{\gamma^{(n)}} )^2 ~.
\label{photon1}
\eeqn
For a photon 
$C(z; \lambda_{\gamma^{(0)}}=0 ) = {\rm const} = \sqrt{\strut r_{\gamma^{(0)}}/L}$.
Note that $s_\phi = \tan \theta_W$ and 
$1/\sqrt{ 1 + s_\phi^2} = \cos \theta_W$.

\bigskip 

\noindent
{\bf (ii) W boson  tower ($\hat W_\mu$)}
\beqn
&&\hskip -1.cm
h_{W^{(n)}}^L = h_{W^{(n)}}^R =  
\frac{1}{\sqrt{2 r_{W^{(n)}}}} \, C(z; \lambda_{W^{(n)}} ) ~, \cr
\noalign{\kern 5pt}
&&\hskip -1.cm 
h_{W^{(n)}}^\wedge = -
\frac{1}{\sqrt{r_{W^{(n)}}}} \, \hat S(z; \lambda_{W^{(n)}} ) ~, \cr
\noalign{\kern 5pt}
&&\hskip -1.cm
r_{W^{(n)}} 
= \int_1^{z_L} \frac{dz}{k z}\,  \Big\{ C(z; \lambda_{W^{(n)}} )^2
+ \hat S (z; \lambda_{W^{(n)}} )^2 \Big\} ~. 
\label{Wboson1}
\eeqn

\bigskip 

\noindent
{\bf (iii) Z boson  tower ($\hat Z_\mu$)}
\beqn
&&\hskip -1.cm
h_{Z^{(n)}}^L = h_{Z^{(n)}}^R = 
\frac{c_\phi^2}{\sqrt{1+ s_\phi^2}}
\frac{1}{\sqrt{\strut 2 r_{Z^{(n)}}}} \, C(z; \lambda_{Z^{(n)}} ) 
= \frac{1- 2 \sin^2 \theta_W}{\cos \theta_W} 
\frac{C(z; \lambda_{Z^{(n)}} )}{\sqrt{2r_{Z^{(n)}} }} ~, \cr
\noalign{\kern 5pt}
&&\hskip -1.cm 
h_{Z^{(n)}}^\wedge = - \sqrt{1+ s_\phi^2}
\frac{1}{\sqrt{ r_{Z^{(n)}}}} \, \hat S(z; \lambda_{Z^{(n)}} ) 
= - \frac{1}{\cos \theta_W} \frac{\hat S(z; \lambda_{Z^{(n)}} )}{\sqrt{r_{Z^{(n)}} }}~, \cr
\noalign{\kern 10pt}
&&\hskip -1.cm
h_{Z^{(n)}}^B  
=-  \frac{\sqrt{2} s_\phi c_\phi}{\sqrt{1 + s_\phi^2}} 
\frac{1}{\sqrt{\strut  r_{Z^{(n)}}}}
C(z; \lambda_{Z^{(n)}} ) 
= - \frac{g_A}{g_B} \frac{\sin^2 \theta_W}{\cos \theta_W} 
\frac{\sqrt{2} C(z; \lambda_{Z^{(n)}} )}{\sqrt{r_{Z^{(n)}} }} ~, \cr
\noalign{\kern 5pt}
&&\hskip -1.cm
r_{Z^{(n)}} = \int_1^{z_L} \frac{dz}{k z}\,  
\Big\{ c_\phi^2 C(z; \lambda_{Z^{(n)}} )^2
+ (1+s_\phi^2) \hat S (z; \lambda_{Z^{(n)}} )^2 \Big\} ~.
\label{Zboson1}
\eeqn

\bigskip 
\noindent
{\bf (iv) $\hat W'_\mu$  tower}
\beqn
&&\hskip -1.cm
h_{W^{\prime (n)}}^L = - h_{W^{\prime (n)}}^R =  \frac{1}{\sqrt{2}} 
\frac{1}{\sqrt{ r_{W^{\prime (n)}}}} \, C(z; \lambda_{W^{\prime (n)}} ) ~, \cr
\noalign{\kern 5pt}
&&\hskip -1.cm
r_{W^{\prime (n)}} 
= \int_1^{z_L} \frac{dz}{k z}\,   C(z; \lambda_{W^{\prime (n)}} )^2  ~.
\label{Wprime1}
\eeqn

\bigskip 
\noindent
{\bf (v) $\hat Z'_\mu$  tower}
\beqn
&&\hskip -1.cm
h_{Z^{\prime (n)}}^L = - h_{Z^{\prime (n)}}^R =  \frac{1}{\sqrt{2}} 
\frac{1}{\sqrt{ r_{Z^{\prime (n)}}}} \, C(z; \lambda_{Z^{\prime (n)}} ) ~, \cr
\noalign{\kern 5pt}
&&\hskip -1.cm
r_{Z^{\prime (n)}} 
= \int_1^{z_L} \frac{dz}{k z}\,   C(z; \lambda_{Z^{\prime (n)}} )^2  ~.
\label{Zprime1}
\eeqn

\bigskip 
\noindent
{\bf (vi) $\hat A^{\hat 4}_\mu$  tower}
\beqn
&&\hskip -1.cm
h_{A^{(n)}} = 
\frac{1}{\sqrt{ r_{A^{\hat 4 (n)}}}} \, S(z; \lambda_{A^{\hat 4 (n)}} ) ~, \cr
\noalign{\kern 5pt}
&&\hskip -1.cm
r_{A^{\hat 4 (n)}} 
= \int_1^{z_L} \frac{dz}{k z}\,   S(z; \lambda_{A^{\hat 4 (n)}} )^2  ~.
\label{Ahat1}
\eeqn

The mode functions for the fifth-dimensional component are given similarly.
$h_{S}^{LR}, h_{D}^{LR}, h_B \propto C'(z; \lambda)$ and
$h_{H}^\wedge, h_{D}^\wedge  \propto S'(z; \lambda)$.
The normalization condition is given by
$\int_1^{z_L}
   (k dz/z) \, (h_{\textrm{\scriptsize s}})^2=1$
where $h_{\textrm{\scriptsize s}}
=h_S^{LR}, h_H^\wedge, h_D^{LR}, h_D^{\wedge}, h_B$.

\subsubsection*{Fermions}

For $P_H$-even $\psi_{{2\over 3}(+)}^{(n)}$,
the equation~(\ref{FermionKK2}) leads to
\beeq
\Big[ \,  a_{U}^{(n)},  a_{t'}^{(n)} \Big] \simeq
 \bigg[  - \frac{\sqrt{2}\tilde{\mu}}{\mu_2} ,
 - \frac{C_L(1;\lambda_n ,c_t)}{S_L(1;\lambda_n , c_t)} \bigg] \, a_{B+t}^{(n)} .
\eneq
With this ratio, the coefficient is given by
\beeq
 a_{B+t}^{(n)} = \bigg[ \int_1^{z_L} dz
   \bigg\{ \Big[2\Big( \frac{\tilde{\mu}}{\mu_2} \Big)^2  +1\Big] 
   C_L(z;\lambda_n ,c_t)^2
+ \hat S_L(z;\lambda_n, c_t)^2   \bigg\} \bigg]^{-1/2} ~.
\label{fermionC1}
\eneq
For $P_H$-odd $t_{(-)}^{(n)}$, the coefficient is
\beeq
a_{B-t}^{(n)} =\bigg[ \int_1^{z_L} dz \,
C_L (z;\lambda_{t_{(-)}^{(n)}}, c_t)^2 \bigg]^{-1/2} ~ .
\label{fermionC2}
\eneq
For $P_H$-even $\psi_{-{1\over 3}(+)}^{(n)}$, the coefficients satisfy
\beeq
\Big[\,  a_{b}^{(n)}, a_{b'}^{(n)} \Big]
\simeq \bigg[  - \frac{\sqrt{2}\mu_2}{\tilde{\mu}} ,
 - \frac{C_L(1;\lambda_n ,c_t)}{S_L(1;\lambda_n , c_t)} \bigg] \,  a_{D+X}^{(n)} ~,
\label{fermionC3}
\eneq
which yields
\beeq
a_{D+X}^{(n)} = \bigg[  \int_1^{z_L} dz \,
\bigg\{ \Big[ 2\Big( \frac{\mu_2}{\tilde{\mu}} \Big)^2 +1\Big]  
C_L(z;\lambda_n ,c_t)^2  + \hat S_L(z; \lambda_n, c_t)^2 \bigg\} \bigg]^{-1/2}  ~.
\label{fermionC4}
\eneq
For $P_H$-odd $b_{(-)}^{(n)}$, the coefficient is
given by
\beeq
a_{D-X}^{(n)} =\bigg[ \int_1^{z_L} dz \,
C_L (z;\lambda_{b_{(-)}^{(n)}}, c_t)^2 \bigg]^{-1/2}  ~.
\label{fermionC5}
\eneq

To obtain  overlap integrals for the gauge couplings, these normalization constants
are taken into account. 
 
\end{appendix}

\vskip 1cm

\renewenvironment{thebibliography}[1]
         {\begin{list}{[$\,$\arabic{enumi}$\,$]}  % {\arabic{enumi}.}
         {\usecounter{enumi}\setlength{\parsep}{0pt}
          \setlength{\itemsep}{0pt}  \renewcommand{\baselinestretch}{1.2}
          \settowidth
         {\labelwidth}{#1 ~ ~}\sloppy}}{\end{list}}

% A useful Journal macro
%\def\jnl#1#2#3#4{{#1}{\bf #2} (#4) #3}
\def\jnl#1#2#3#4{{#1}{\bf #2},  #3 (#4)}

\def\Zphys{{\em Z.\ Phys.} }
\def\jssc{{\em J.\ Solid State Chem.\ }}
\def\jpsJ{{\em J.\ Phys.\ Soc.\ Japan }}
\def\ptps{{\em Prog.\ Theoret.\ Phys.\ Suppl.\ }}
\def\PTP{{\em Prog.\ Theoret.\ Phys.\  }}
\def\JMP{{\em J. Math.\ Phys.} }
\def\NPB{{\em Nucl.\ Phys.} B}
\def\NP{{\em Nucl.\ Phys.} }
\def\PLB{{\it Phys.\ Lett.} B}
\def\PL{{\em Phys.\ Lett.} }
\def\PRL{\em Phys.\ Rev.\ Lett. }
\def\PRB{{\em Phys.\ Rev.} B}
\def\PRD{{\em Phys.\ Rev.} D}
\def\PRe{{\em Phys.\ Rep.} }
\def\AP{{\em Ann.\ Phys.\ (N.Y.)} }
\def\RMP{{\em Rev.\ Mod.\ Phys.} }
\def\ZPC{{\em Z.\ Phys.} C}
\def\SCI{\em Science}
\def\CMP{\em Comm.\ Math.\ Phys. }
\def\MPLA{{\em Mod.\ Phys.\ Lett.} A}
\def\IJMPA{{\em Int.\ J.\ Mod.\ Phys.} A}
\def\IJMPB{{\em Int.\ J.\ Mod.\ Phys.} B}
\def\EPJC{{\em Eur.\ Phys.\ J.} C}
\def\PR{{\em Phys.\ Rev.} }
\def\JHEP{{\em JHEP} }
\def\JCAP{{\em JCAP} }
\def\cmp{{\em Com.\ Math.\ Phys.}}
\def\JPA{{\em J.\  Phys.} A}
\def\JPG{{\em J.\  Phys.} G}
\def\NJP{{\em New.\ J.\  Phys.} }
\def\CQG{\em Class.\ Quant.\ Grav. }
\def\ATMP{{\em Adv.\ Theoret.\ Math.\ Phys.} }
\def\ibid{{\em ibid.} }

\renewenvironment{thebibliography}[1]
         {\begin{list}{[$\,$\arabic{enumi}$\,$]}  % {\arabic{enumi}.}
         {\usecounter{enumi}\setlength{\parsep}{0pt}
          \setlength{\itemsep}{0pt}  \renewcommand{\baselinestretch}{1.2}
          \settowidth
         {\labelwidth}{#1 ~ ~}\sloppy}}{\end{list}}

%%%%  title of reference  %%%%
\def\reftitle#1{{\it ``#1'' }}    %to print.
%\def\reftitle#1{}                %to hide.

%%%%%%%%%%%%% BIBLIOGRAPHY (US) %%%%%%%%%%%%%%%%%%%%

\end{document}